\documentclass[twocolumn]{aastex631}

\newcommand\asloth{\textsc{a-sloth}}
\newcommand{\Msun}{\,\ensuremath{\mathrm{M}_\odot}}
\newcommand{\Rv}{\ensuremath{R_\mathrm{vir}}}
\newcommand{\mvir}{\ensuremath{M_\mathrm{vir}}}
\newcommand{\mpeak}{\ensuremath{M_\mathrm{vir,peak}}}
\newcommand{\mste}{\ensuremath{M_*}}
\newcommand{\te}{\ensuremath{\tau_e}}

\usepackage{amsmath}
\usepackage{xcolor}

\usepackage{xspace}
\newcommand{\ctp}{Caterpillar\xspace}
\usepackage{lipsum}

\shorttitle{Public release of A-SLOTH}
\shortauthors{Hartwig et al.}

\graphicspath{{./}{./}}

\begin{document}

\title{Public Release of A-SLOTH: Ancient Stars and Local Observables by Tracing Halos}

\newcommand{\SoS}{Department of Physics, School of Science, The University of Tokyo, Bunkyo, Tokyo 113-0033, Japan}
\newcommand{\ipi}{Institute for Physics of Intelligence, School of Science, The University of Tokyo, Bunkyo, Tokyo 113-0033, Japan}
\newcommand{\IPMU}{Kavli Institute for the Physics and Mathematics of the Universe (WPI), The University of Tokyo Institutes for Advanced Study, The University of Tokyo, Kashiwa, Chiba 277-8583, Japan}
\newcommand{\ITA}{Universit\"at Heidelberg, Zentrum f\"ur Astronomie, Institut f\"ur Theoretische Astrophysik, Albert-Ueberle-Str.\ 2, 69120 Heidelberg, Germany}
\newcommand{\imprs}{International Max Planck Research School for Astronomy and Cosmic Physics at the University of Heidelberg (IMPRS-HD), \\ Königstuhl 17, D-69117 Heidelberg, Germany}

\correspondingauthor{Tilman Hartwig}
\email{hartwig@phys.s.u-tokyo.ac.jp}

\author[0000-0001-6742-8843]{Tilman Hartwig}
\affiliation{\ipi}
\affiliation{\SoS}
\affiliation{\IPMU}

\author[0000-0002-9022-5136]{Mattis Magg}
\affiliation{\ITA}
\affiliation{\imprs}

\author[0000-0003-0475-1947]{Li-Hsin Chen}
\affiliation{\ITA}
\affiliation{\imprs}

\author[0000-0002-9801-7788]{Yuta Tarumi}
\affiliation{\SoS}

\author[0000-0003-0212-2979]{Volker Bromm}
\affiliation{Department of Astronomy, University of Texas at Austin, Austin, TX 78712, USA}

\author[0000-0001-6708-1317]{Simon C. O. Glover}
\affiliation{\ITA}

\author[0000-0002-4863-8842]{Alexander P. Ji}
\affiliation{Department of Astronomy \& Astrophysics, University of Chicago, 5640 S Ellis Avenue, Chicago, IL 60637, USA}
\affiliation{Kavli Institute for Cosmological Physics, University of Chicago, Chicago, IL 60637, USA}

\author[0000-0002-0560-3172]{Ralf S. Klessen}
\affiliation{\ITA}
\affiliation{Universit\"{a}t Heidelberg, Interdisziplin\"{a}res Zentrum f\"{u}r Wissenschaftliches Rechnen, Im Neuenheimer Feld 225, 69120 Heidelberg, Germany}

\author[0000-0003-2480-0988]{Muhammad A. Latif}
\affiliation{Physics Department, College of Science, United Arab Emirates University, PO Box 15551, Al-Ain, UAE}

\author[0000-0002-3216-1322]{Marta Volonteri}
\affiliation{Institut d'Astrophysique de Paris, CNRS \& Sorbonne Université, UMR 7095, 98 bis bd Arago, F-75014 Paris, France}

\author[0000-0001-7925-238X]{Naoki Yoshida}
\affiliation{\ipi}
\affiliation{\SoS}
\affiliation{\IPMU}

\begin{abstract}
The semi-analytical model \asloth\ (Ancient Stars and Local Observables by Tracing Halos) is the first public code that connects the formation of the first stars and galaxies to observables. After several successful projects with this model, we publish the source code\footnote{\url{https://gitlab.com/thartwig/asloth}} and describe the public version in this paper.
The model is based on dark matter merger trees that can either be generated based on Extended Press-Schechter theory or that can be imported from dark matter simulations. On top of these merger trees, \asloth\ applies analytical recipes for baryonic physics to model the formation of both metal-free and metal-poor stars and the transition between them with unprecedented precision and fidelity. \asloth\ samples individual stars and includes radiative, chemical, and mechanical feedback. It is calibrated based on six observables, such as the optical depth to Thomson scattering, the stellar mass of the Milky Way and its satellite galaxies, the number of extremely-metal poor stars, and the cosmic star formation rate density at high redshift. \asloth\ has versatile applications with moderate computational requirements. It can be used to constrain the properties of the first stars and high-z galaxies based on local observables, predicts properties of the oldest and most metal-poor stars in the Milky Way, can serve as a subgrid model for larger cosmological simulations, and predicts next-generation observables of the early Universe, such as supernova rates or gravitational wave events.
\end{abstract}



\section{Introduction} \label{sec:intro}
One of the main goals of astrophysics is to understand the formation of stars and galaxies in order to explain the evolution of the Universe from an almost homogeneous state shortly after the Big Bang until the present day. There are many open questions on small and large scales about the formation of stars and evolution of galaxies. One possible approach to answer these questions is to model the temporal evolution of the Universe from high redshift to $z=0$ \citep[e.g.,][]{peebles93}.

Structures in the Universe formed hierarchically under the influence of gravity. Small overdensities in the early Universe evolved into the first gravitationally bound structures, which then hosted the first stars. These  small early structures then merged with other halos to form bigger structures, which eventually became the galaxies that we observe in the local Universe. In general, the evolution of the Universe proceeds from simple to complex, from small to big, from pristine to chemically enriched.

While we have many observations of phenomena in the local Universe, our understanding of the cosmos decreases as we move further back in time. For example, we understand star formation in the Milky Way (MW) sufficiently well to model the mass distribution of newly-formed stars or predict the final fates of nearby stars. However, we have only a vague understanding of the formation of the first generations of stars due to a lack of direct observables \citep{Glover05,bromm11,greif15,klessen19}.

Numerical simulations and semi-analytical models (SAMs) can bridge this gap. Starting from ab-initio initial conditions, one can model the successive formation of stars and galaxies in time and eventually compare the results of such models with observations of the real Universe. However, numerical simulations have to compromise between mass resolution and effective volume or use a zoom-in approach in which only some part of the volume is simulated with higher resolution. For example, TNG50 \citep{Nelson2019,Pillepich2019}, the highest resolution cosmological simulation in the IllustrisTNG project, has a mass resolution of $m_\mathrm{DM} = 4.5\times10^5\Msun$ and $m_\mathrm{baryon} = 8.5\times10^4\Msun$ for dark matter (DM) and baryons, respectively. Therefore, the smallest galaxies reliably identified in TNG50 \citep{Engler2021} have stellar masses of $\sim 6.5\times10^6\Msun$ or more, far larger than the values we expect to find in the first minihalos at high redshift. On the other hand, numerical simulations with high enough resolution to reliably model individual minihalos \citep[e.g.][]{Schauer19a} typically involve volumes that are too small to allow them to follow the formation of massive galaxies at low redshift.

SAMs are computationally more efficient (see below). Depending on their specific flavor, they approximate galaxies with a few characteristic parameters and components and follow their evolution through time. This makes SAMs computationally very efficient and allows one to focus on a few specific aspects of the model. For example, one can trace certain types of stars from high redshift to the present-day MW.

Moreover, SAMs allow to explore a broad parameter space due to their modest runtime. While large cosmological simulations have runtimes of several million CPU hours on modern supercomputers \citep{Pillepich2019,yoshikawa21}, a SAM that covers a similar effective volume can run on an ordinary computer within a few minutes.

This computational efficiency makes SAMs ideally suited to study poorly constrained phenomena in the early Universe that have an observable effect in the local Universe. For example, the formation of the first stars in the Universe (Population~III or Pop~III) is not well understood. However, it affects the epoch of reionization \citep{bromm11,dayal18}, the formation of the first supermassive black holes \citep[SMBHs,][]{latif16,woods19,inayoshi20}, and the population of old and metal-poor stars in the MW and its satellites \citep{frebel15,ishigaki21}. A SAM allows us to model the formation of the first stars under different assumptions (e.g. regarding their mass distribution or star formation efficiency) and determine which assumption agrees best with observations, and the models that reproduce successfully known observables can be used to predict new diagnostics. Therefore, SAMs provide a fruitful connection between the high-redshift Universe and local observables.

In the next section, we first present our previous models that lead to the development of \asloth, then discuss and compare various other SAMs in the literature, and eventually present the benefits of \asloth.

\subsection{A-SLOTH Ancestors}
\label{sec:PreviousSloth}
\begin{figure}[ht]
\centering
\includegraphics[width=0.66\columnwidth]{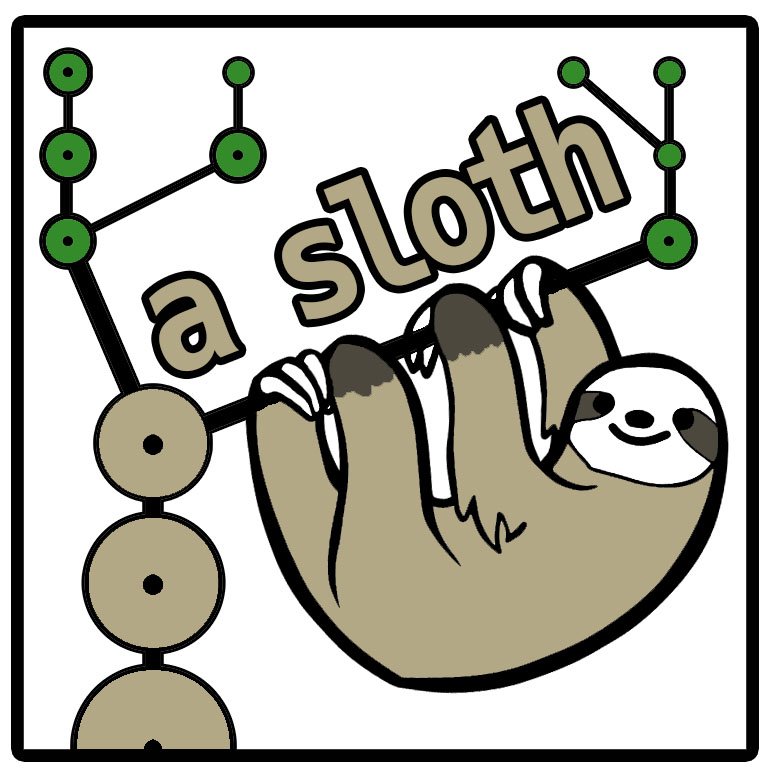}
\caption{\asloth\ conquering a merger tree.  \label{fig:logo}}
\end{figure}
\asloth\ (Fig.~\ref{fig:logo}) was originally developed to model the formation and distribution of possible low-mass Pop~III survivors in the MW \citep{Hartwig15b} and is based on \textsc{Galform} \citep{Parkinson2008}. We then extended \asloth\ to be applicable for a variety of questions relating to the high-redshift Universe as well as the oldest stars in the MW. Among the topics we examined were whether the Lyman-$\alpha$ emitter CR7 \citep{Sobral15} could be a Pop~III galaxy \citep{Hartwig16b}, surviving metal-free stars in MW satellites \citep{Magg18}, constraints on the pristine initial mass function (IMF) \citep{Tarumi20} and the low-mass end of the satellite mass function of the MW \citep{chen22}. Each of these works built upon the previous models and contributed to the development of \asloth.

We also implemented a weighting of merger trees based on their mass and predicted the rates of SNe from the first stars \citep{magg16}. This weighting allows us to sample halo masses from the halo mass function in order to create a cosmologically representative effective volume from merger trees generated using the Extended Press-Schechter (EPS) formalism (see Section~\ref{sec:eps} below). We also applied this updated model to predict the gravitational wave (GW) signals from the remnants of the first stars and showed that the detection rate of Pop~III GW events is lower than events from other channels, but that Pop~III remnants could be identified in the future by their higher masses \citep{hartwig16a}. In an independent study, \citet{liu21} adapted \asloth\ to study GWs from the remnants of the first stars in nuclear star clusters and found promising detection rates.

Our improved treatment of metallicity allowed us to model the transition from Pop~III to Pop~II star formation more realistically, and we have implemented a new model for inhomogeneous metal mixing in the first galaxies \citep{hartwig18a,Tarumi20,ishigaki21}. This allows us to predict the properties of second-generation and extremely metal-poor (EMP) stars.

An updated model of star formation allowed us to sample individual Pop~II stars and to simulate the formation of MW satellites \citep{chen22}. \asloth\ can also be used as a subgrid model to larger cosmological simulations, e.g., to predict the impact of the first generations of stars on the global 21cm signal \citep{Magg21b}.

\subsection{Other Semi-Analytical Models}
SAMs are popular in astrophysics due to their moderate computational costs. They approximate and simplify physical processes to model the Universe, and each has been developed for a specific purpose which limits their applicability. Here, we summarize SAMs in the literature, discuss their strengths, and later show how \asloth\ improves upon several aspects of these existing models. We limit this discussion to only SAMs that model Pop~III star formation explicitly. There are further notable SAMs that model the pre-reionization Universe but that do not focus on the formation of the first generations of stars \citep{Volonteri03,barausse12,starkenburg13,agarwal16,pacucci18}. A comparison of further SAMs is presented in \citet{knebe15}.

\textsc{gamete} \citep{Salvadori07} and its extension \textsc{gamete/qsodust} \citep{valiante16,Valiante18} model early star and galaxy formation based on EPS-generated merger trees. The model is suited for all redshifts. It has an arbitrarily fine mass resolution and accounts for radiative, chemical, and mechanical feedback. To distinguish between Pop~III and Pop~II star formation, the authors use a criterion based on a critical metallicity $Z_\mathrm{crit}$ \citep{BrommLoeb03}, i.e., on the total metal and dust content of a halo.

They calibrate their model by reproducing global properties of the Milky Way at z=0 and the MDF of Galactic halo stars \citep{Salvadori07}. The model is also able to reproduce the observed properties of Local Group dwarf galaxies \citep[e.g.][]{salvadori15} and, by modeling the inhomogeneous reionization and metal-enrichment, it has been used to study high-z Damped Lyman Alpha absorption systems \citep{salvadori12}, dwarf galaxies as reionization sources \citep{salvadori14}, and the IMF of the first black hole seeds \citep{ferrara14}. The latest improvement of the model also accounts for the random sampling of the Pop~III IMF \citep{deBennassuti17}. They are also able to reproduce a sample of quasars at high redshift \citep{valiante14} with a subgrid model for the seeding, growth, and feedback of SMBHs \citep{valiante16,Valiante18,sassano21,trinca22}. The \textsc{gamete} code has also been combined with N-body simulations to gain spatial information \citep{salvadori10,pacucci17}.


\citet{graziani17} apply the model \textsc{gamesh} \citep{graziani15} to simulate the formation of a MW-like halo. They follow the baryonic evolution and radiative transfer in a semi-analytic way. This model reproduces the observed galaxy main sequence, the mass-metallicity and the Fundamental Plane of metallicity relations in the redshift range $z=0-4$.
    
The SAM presented in \citet{visbal20} simulates the formation of the first stars based on merger trees from $N$-body simulations. It has a mass resolution of $\sim 10^5\Msun$ and operates in the redshift range $z=6-30$. The authors include feedback in the form of ionizing radiation, Lyman-Werner (LW) photons, and metals based on a computationally efficient fast Fourier transform. They also take into account baryonic streaming and reproduce the galaxy UV luminosity function at high redshift.
    
\textsc{delphi} \citep{dayal14,dayal17,dayal19,dayal20,chatterjee20} is a versatile model used to simulate the formation of galaxies and SMBHs at high redshift. It operates down to $z=4$ and allows for different cosmologies (other than $\Lambda$CDM). The authors include external radiative feedback and can reproduce various independent observables, such as the UV luminosity function of galaxies and quasars at $z=6$, star formation rate density (SFRd), the black hole mass function, and the black hole mass -- stellar mass relation.
    
\textsc{gamma} is a SAM used to probe the impact of nuclear astrophysics uncertainties on galactic chemical evolution and to address the chemical evolution of dwarf galaxies in the early universe \citep{cote17,cote18}. It uses spatially resolved DM merger trees from $N$-body simulations and operates down to $z \sim 7$. It includes inhomogeneous metal mixing in post-processing to reproduce the metallicity distribution function (MDF) of the Renaissance simulation \citep{ren15}. \textsc{gamma} traces individual elements and even s- and r-process elements.
    
\textsc{startree} models the formation of the first and second generation of stars to constrain properties of the first stars based on EMP stars in the MW. It is based on EPS trees and follows individual Pop~III stars \citep{komiya09,komiya10}. The model includes LW and ionizing feedback, a stochastic model for the inhomogeneous distribution of metal enrichment inside a proto-galaxy, faint Pop~III SNe, and allows the user to trace individual elements. The authors also take into account the possible surface pollution of low-mass Pop~III stars from accreted metal-enriched interstellar medium (ISM) \citep{komiya10,Komiya16}. The model reproduces the observed mass–metallicity relation for dwarf galaxies and various abundance ratio distributions. It operates in the redshift range $z=0-30$ and can also be used to model the formation of EMP \citep{komiya11}, and CEMP stars \citep{komiya20}.

The SAM by \citet{tumlinson06} models the chemical evolution of MW-like galaxies to constrain the properties of the first and second generation of stars. The original version is based on EPS-generated merger trees, and an updated model uses merger trees from $N$-body simulations as input \citep{tumlinson10,gomez14}. The authors include chemical, radiative, and mechanical feedback. The model is calibrated to reproduce the MDF, ionization history, relative elemental abundances of EMP stars, the MW stellar mass, and the satellite luminosity function.

\citet{Trenti09} developed a SAM based on EPS merger trees to model the formation of the first and second generation of stars. The model includes radiative feedback and stochastic metal enrichment. The authors do not constrain their model based on observables but instead vary various input parameters to study the effect of such parameters on the model.

Finally, the SAM by \citet{Crosby13} is based on $N$-body simulations and allows the user to study cosmic variance. It includes radiative, chemical, and mechanical feedback. The authors constrain their model based on the SFRd at $z = 10.3$.

\subsection{Advantages of \asloth}
\asloth\ is the first public SAM that models the formation of the first stars and galaxies and connects those to local observables. It is calibrated transparently, which guarantees reproducibility and has several advantages over existing SAMs.

\asloth\ uses both EPS-based merger trees and merger trees extracted from $N$-body simulations as input. We provide instructions to obtain spatially resolved merger trees, or users can input their own data. Currently, \asloth\ supports merger trees in the format of \textsc{consistent-trees} \citep{ConsistentTrees}, but the input routine can be modified easily. Other SAMs that use only EPS-generated merger trees \citep[e.g.][]{Salvadori07,komiya09,dayal14} lack spatial information for the feedback model.

\asloth\ is highly efficient and parallelized. We have implemented a tree-based lookup for external feedback and minimized the memory requirements as much as practically possible. A MW-like galaxy can be simulated in just over 1 minute on 4 cores, and a cosmologically representative box with side length 8\,Mpc/h can be simulated down to redshift $z=4.5$ on 32 cores in only 14 minutes. The SAM by \citet{visbal20} evaluates feedback with a computationally efficient fast Fourier transform. Other SAMs do not comment on their runtime efficiency. For more details on the numerical implementation of \asloth, see Magg et al., in prep.

\asloth\ is the only SAM that samples and traces individual Pop~III and Pop~II stars. The stochastic sampling of individual stars and their SNe is essential to correctly model feedback in the first galaxies \citep{chen22} and the tracking of individual Pop~II stars allows us to directly compare our results to metal-poor stars in the MW. Only few EPS-based models \citep{tumlinson06,komiya09,deBennassuti17} sample individual Pop~III stars. In the nomenclature of \asloth, we distinguish between Pop~III (metal-free) and Pop~II (metal-poor) stars. The group of Pop~II stars in \asloth\ also includes Pop~I (about Solar metallicity) stars, but the exact distinction between Pop~II and Pop~I is not relevant for all current applications of \asloth.

\asloth\ tracks individual elements. In the current version, we track carbon and iron explicitly, which allows us to apply a more sophisticated, observationally motivated criterion to distinguish between Pop~III and Pop~II star formation \citep{Chiaki18}. Some SAMs also track individual elements \citep{cote18,komiya20} or even include dust \citep{valiante14,valiante16,deBennassuti17} while others use a simple metallicity cut to discriminate between Pop III and Pop II stars \citep{tumlinson06}.

\asloth\ is calibrated based on six observables, and our calibration procedure is robust and transparent. SAMs usually have various free parameters that need to be chosen. Depending on the application, one guesstimates or calibrates these input parameters to reproduce specific data. Other SAMs aim to reproduce the MDF \citep{tumlinson06,komiya10}, observations of quasars or the UV luminosity function at high redshift \citep{dayal14,Valiante18,visbal18,trinca22}, properties of the MW and its dwarf galaxies \citep{salvadori15,deBennassuti17}, or results from simulations \citep{cote18}. \asloth\ is calibrated to reproduce observables from both the local and high-redshift Universe simultaneously, and we show that this calibration also reproduces additional constraints that we did not actively aim for, such as the metal-enriched volume fraction, the recovery time distribution from simulations, or the observed stellar-mass-to-halo-mass (SMHM) relation.

\asloth\ models the recovery time \citep{Jeon14} between Pop~III and Pop~II star formation self-consistently by following the baryonic contents over time. Most models assume a constant recovery time, which defines the time between the last Pop~III SN explosion and the first Pop~II star formation in a halo. We solve the differential equations that govern the heating and cooling of gas inside minihalos in order to calculate when there is enough gas re-accreted after the first SNe to trigger second-generation star formation.

\asloth\ includes a subgrid recipe for inhomogeneous metal mixing inside the first galaxies, which is crucial to correctly model the metallicity of second-generation stars \citep{Hartwig19,Tarumi20}. Only the SAM of \citet{komiya20} applies a similar model that takes into account inhomogeneous mixing of metals. \citet{salvadori12} account for incomplete mixing of metals in SN outflows. To mimic the effect of inhomogeneous mixing, \citet{cote18} apply a smoothing kernel to their MDF in post-processing in order to match results from simulations.

\section{Methodology}
\subsection{Input Data}
Throughout the code, we assume a flat $\Lambda$CDM Universe and use cosmological parameters from \citet{Planck2015}. Specifically, we use the Hubble constant $H_0 = 68\,\mathrm{km}\,\mathrm{s}^{-1}\,\mathrm{Mpc}^{-1} = 100h\,\mathrm{km}\,\mathrm{s}^{-1}\,\mathrm{Mpc}^{-1}$, a matter density of $\Omega_m = 0.31$, and a baryon mass fraction of $\Omega_b = 0.049$.

The primary input to \asloth\ is a DM halo merger tree describing the merger history of the DM halos whose stellar and gas content we wish to model. These merger trees can be generated using the EPS formalism or extracted from a cosmological simulation, as described in more detail below. 

\subsubsection{Extended Press-Schechter Merger Trees}
\label{sec:eps}
The ``Press-Schechter'' formalism \citep{PressSchechter} provides an analytical estimate for the number density of DM halos as a function of their mass and redshift. The comoving number density of halos of mass between $M$ and $M+\mathrm{d}M$ is given by
\begin{equation}
 \frac{\mathrm{d}n}{\mathrm{d}M} = \sqrt{\frac{2}{\pi}} \frac{\rho _m}{M} \frac{-\mathrm{d}(\ln \sigma (M))}{\mathrm{d} M} \nu _c \exp (- \nu _c ^2 /2),  \label{eq:ps}
 \end{equation}
where $\rho _m$ is the matter density and $\sigma (M)$ is the standard deviation of the matter power spectrum smoothed on a mass scale $M$. The parameter $\nu _c = \delta _c (z)/ \sigma (M)$ is the critical threshold for collapse, where $\delta _c$ is the overdensity for non-linear collapse with $\delta _c(z=0)=1.686$.

The original derivation of Equation~\ref{eq:ps} by \citet{PressSchechter} suffered from a normalization problem: the mass function they derived accounted for only half the mass in the Universe, with Equation~\ref{eq:ps} following only after they multiplied their original expression by a factor of two, a procedure with little justification in their study. Later, \citet{Bond1991} showed that this normalization problem is a consequence of the cloud-in-cloud problem, i.e.\ the difficulty of correctly counting clouds in hierarchical models where smaller clouds can be embedded within larger ones. They showed that this problem could be avoided by an approach based on excursion sets of the density field. In this approach, the DM density field is smoothed on a series of successively smaller scales $R$, each of which is associated with a corresponding mass scale $M(R)$. Points in the density field that first exceed the overdensity $\delta_c$ when smoothed on scale $R$ are then associated with halos of mass $M$. In the simple case where the smoothing function is a sharp ``top-hat'' filter in wave-number space, the mass function that one obtains from this approach is identical to Equation~\ref{eq:ps}, i.e.\ the method avoids the original normalization problem. 

\citet{LaceyCole1993} subsequently extended this idea by showing that the excursion set approach allows one to derive not only the DM halo mass function but also the halo merger rate, as well as specific merger histories for individual halos. In \asloth, we use the publicly available implementation of this EPS formalism presented by \citet{Parkinson2008}, which is based on the code by \citet{cole00}.

The EPS formalism as used in the original \textsc{galform} code \citep{cole00} systematically under-predicts the mass of the most massive progenitors at higher redshifts. Hence, we use the updated version of the code by \citet{Parkinson2008}, which modifies the progenitor mass function with a perturbing function
\begin{equation}
 \frac{\mathrm{d}N}{\mathrm{d} M_1} \rightarrow \frac{\mathrm{d}N}{\mathrm{d} M_1} G(\sigma _1 / \sigma _2 , \delta _2 / \sigma _2)
\end{equation}
to reproduce the halo merger histories of the Millennium simulation \citep{springel05}. The best-fitting perturbing function is given by
\begin{equation}
 G(\sigma _1 / \sigma _2 , \delta _2 / \sigma _2) = 0.57 \left( \frac{\sigma _1}{\sigma _2} \right) ^{0.38} \left( \frac{\delta _2}{\sigma _2} \right) ^{-0.01}.
\end{equation}
We have chosen this specific implementation of the merger tree, because it performs best compared to other EPS implementations \citep{Jiang2014}.

\subsubsection{Merger Trees from N-Body Simulations}
In addition to statistically generated merger trees, we use merger trees from cosmological simulations. In this study we use merger trees from \citet{Ishiyama16} and from the \ctp project \citep{Caterpillar, Griffen18}. These $N$-body DM only simulations both have sufficient resolution to resolve minihalos. For both simulations, halos were obtained with \textsc{Rockstar} \citep{rockstar} and merger trees were generated with \textsc{Consistent-Trees} \citep{ConsistentTrees}. From the \ctp simulations we use merger trees of 30 zoom-in simulations of MW-mass galaxies ($M_{\rm halo} \sim 10^{12} M_\odot$), selected from a (100$h^{-1}$ Mpc)$^3$ parent box. The high-resolution DM particle mass is $3\times10^4 M_\odot$, and the simulations are run to $z=0$ with a high-resolution region that extends at least to the virial radius of each halo. This is the largest suite of high-resolution zoom-in simulations of MW-mass that can resolve the formation of low-mass minihalos, providing a good estimate of cosmic variance of $z=0$ properties related to the first stars and galaxies. They are ideal for applications related to stellar archaeology. The data from \citet{Ishiyama16} include merger trees for every halo within an $(8\,h^{-1}\,\mathrm{Mpc})^{3}$ comoving volume down to a redshift of $z=4$ with a mass resolution of $5000h^{-1}\Msun$. These merger trees allow us to model a large and well-defined cosmological volume for applications such as ionization histories, cosmic star formation histories or the large-scale enrichment of the intergalactic medium (IGM).

Both the \ctp trees and the merger trees from \citet{Ishiyama16} are not one individual tree but rather contain a larger number of separate merger trees. In order to allow for easier processing, we resort the merger trees into a strictly time-ordered manner and save them as a binary file.

For brevity, we refer later to the cosmologically representative merger trees from the box with side length $8\,h^{-1}\,\mathrm{Mpc}$ as the `8\,Mpc box'. For comparison, each set of \ctp merger trees is extracted from a volume with an effective side length of (3 -- 6)\,$h^{-1}\,\mathrm{Mpc}$.

In contrast to the Press-Schechter merger trees, trees from N-Body simulations generally contain subhalos. We prevent accretion of hot gas onto subhalos, because their masses tend to fluctuate significantly during mergers. Otherwise subhalos are treated, followed, and modeled just as normal halos.

\subsubsection{Auxiliary Input Data}
SAMs do not simulate star formation from first principles, such as hydrodynamic simulations do. Instead, \asloth\ relies on analytical formulae and pre-computed input quantities. In this section, we explain and justify our input data and explain how users can modify these assumptions for their own models.
In the following equations, $x = \mathrm{log}_{10} (M_\mathrm{star}/\Msun)$, where $M_\mathrm{star}$ is the zero-age-main-sequence mass of an individual star.
\begin{itemize}
    \item \textbf{Initial mass function} \\
    For Pop~III stars, we assume a power-law IMF of the form
    \begin{equation}
        \frac{dN}{d\mathrm{log}M_\mathrm{star}} \propto M_\mathrm{star}^{-\alpha_{\rm III}}
    \end{equation}
    in the mass range $M_\mathrm{min} \leq M_\mathrm{star}/\Msun \leq M_\mathrm{max}$, where $\alpha_{\rm III}$ is an adjustable parameter.
    For Pop~II stars, we follow the \citet{KroupaIMF} IMF in the mass range (0.01, 100)\Msun\ with the functional form 
    \begin{equation}
    \begin{aligned}
        &\frac{dN}{d\mathrm{log}M_\mathrm{star}} \propto M_\mathrm{star}^{0.7} \mathrm{~for~}  \frac{M_\mathrm{star}}{\Msun} < 0.08 \\
        &\frac{dN}{d\mathrm{log}M_\mathrm{star}} \propto M_\mathrm{star}^{-0.3} \mathrm{~for~} 0.08 \leq \frac{M_\mathrm{star}}{\Msun} < 0.5 \\
        & \frac{dN}{d\mathrm{log}M_\mathrm{star}} \propto M_\mathrm{star}^{-1.3} \mathrm{~for~} 0.5 \leq \frac{M_\mathrm{star}}{\Msun}. \\
    \end{aligned}
    \end{equation}
    \item \textbf{Stellar lifetime} \\
    We compute the stellar lifetimes of Pop III stars with the fitting function \citep{Schaerer2002}
    \begin{equation}
        \mathrm{log}_{10} (T_\mathrm{III}/\mathrm{yr}) = 9.785 - 3.759x + 1.413x^2 - 0.186x^3, \label{eq:tstarIII}
    \end{equation}
    and the stellar lifetimes of Pop~II (1/50 $Z_\odot$) stars with the fitting function \citep{Stahler05}
    \begin{equation}
        \mathrm{log}_{10} (T_\mathrm{II}/\mathrm{yr}) = 10 - 3.68x + 1.17x^2 - 0.12x^3.
    \end{equation}
    These fitting functions were constructed to express the lifetimes of stars in the mass ranges $5-500\Msun$ for Pop~III stars and $7-150\Msun$ for Pop~II stars. However, the values predicted by Eq.~\ref{eq:tstarIII} agree well with the individual stellar lifetimes of Pop~III stars calculated by \citet{Marigo2001} down to sub-solar masses. We therefore use this fitting function down to Pop~III stellar masses of $0.7\Msun$. For lower stellar masses, the lifetimes exceed the current age of the Universe and so the precise values are unimportant.  For Pop~II stars, our adopted fitting function agrees well with data from \citet{Ekstrom08} and \citet{Stahler05} for stellar masses ranging from sub-solar values up to $200\Msun$. Therefore, we also use the fit of stellar lifetimes in this extended range.
    \item \textbf{Ionizing photon production} \\
    The time-averaged ionizing photon production rates of Pop~III and Pop~II stars are computed with the formulae\footnote{These formulae are for single stars. We do not currently account for the impact of binaries on the ionizing photon production rate. However, we expect that their impact will be relatively small. At early times, when photoionization is the most important form of feedback, the ionizing photon production rate is insensitive to the binary fraction, while at later times, when the sensitivity is much larger \citep{Secunda2020}, supernovae will typically be the most important form of feedback.} \citep{Schaerer2002}
    \begin{equation}
        \mathrm{log}_{10} (Q_\mathrm{III}/\mathrm{s}^{-1}) = 43.61 + 4.9x - 0.83x^2, 
    \end{equation}
    and 
    \begin{equation}
        \mathrm{log}_{10} (Q_\mathrm{II}/\mathrm{s}^{-1}) = 27.8 + 30.68x - 14.8x^2 + 2.5x^3,
    \end{equation}
    respectively. These fitting functions are valid in the mass ranges $9-500\Msun$ for Pop~III and $7-150\Msun$ for Pop~II stars.
    We consider all photons with energies above $E_\gamma =13.6\,\mathrm{eV}$ to be ionizing and neglect the ionization of helium.
    \item \textbf{Metal yields ejected by supernovae} \\
    For the metal yields ejected by supernovae, we use tabulated values from \citet{Nomoto13} and \citet{Kobayashi06} for Pop~III and Pop~II stars, respectively. In its current version, \asloth\ follows carbon, iron, and the sum of all elements explicitly. The code can be modified easily to track more or less elements.
\end{itemize}
    
\begin{figure*}
      \centering
      \includegraphics[width=\textwidth]{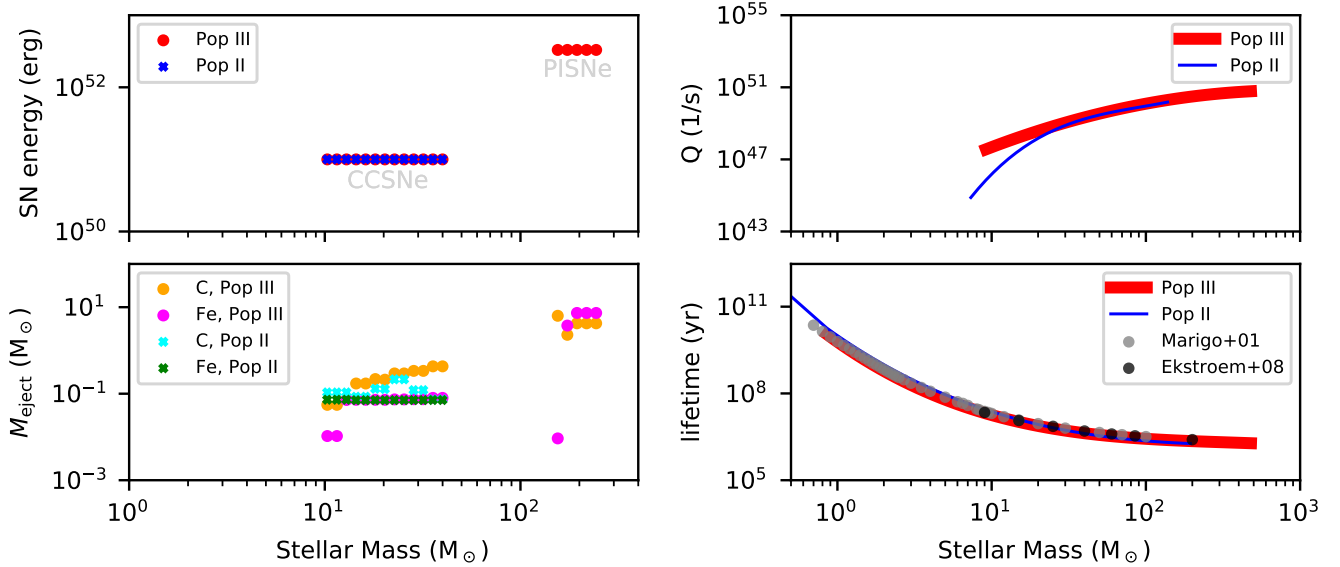}
      \caption{Illustration of SN energy, ionizing photon production rate, carbon and iron ejecta mass, and stellar lifetimes that are available as input data in \asloth. Pop~III stars are plotted in red and Pop~II stars are plotted in blue. Users can define their own specific mass ranges or use other tabulated input data.}
      \label{fig:stellarprop}
\end{figure*}

The implemented SN energy, metal yields, ionizing flux, and stellar lifetimes are illustrated as a function of stellar mass in Fig.~\ref{fig:stellarprop}.
We illustrate the quantities in the stellar mass ranges for which they are valid. However, please note that we sample Pop~II stars only up to $100\Msun$ and Pop~III stars only in the mass range $M_\mathrm{min}=5\Msun$ to $M_\mathrm{max}=210\Msun$ in our fiducial model. Users of \asloth\ can adapt these mass ranges as desired.

We interpolate values onto our grid of stellar masses for lifetime and ionizing flux. For the metal yields, we use the nearest-neighbor values, i.e. the nearest stellar mass available in the tabulated grid of SN yields, so that we do not create artificially mixed combinations by interpolating supernova yields from different progenitor masses. We assume that Pop~II and Pop~III stars in the mass range $10-40\Msun$ explode as core-collapse supernovae (CCSNe), Pop~III stars in the mass range $140-260\Msun$ explode as pair instability supernovae (PISNe) \citep{HegerWoosley2002,HegerWoosley2010}, and that there are no supernova explosions for stellar masses between these two ranges.

\subsection{Baryon Cycle}
\label{sec:bc}
\begin{table}
\centering
\begin{tabular}{ll}
    & halo-specific masses\\
    \hline
    $\mvir$ & virial mass of the halo \\
    $\mpeak$ & peak virial mass of the halo \\
    $M_\mathrm{cold}$ & cold gas mass \\
    $M_\mathrm{hot}$ & hot gas mass \\
    $M_\mathrm{out}$ & cumulative outflow mass \\
    $M_\mathrm{*}$ & total stellar mass \\ 
    $M_\mathrm{disk}$ & disk mass, including gas and stars \\
    $\delta M_\mathrm{out}$ & outflow mass \\
    $\delta M_\mathrm{out, cold}$ & cold gas mass that enters outflow \\
    $\delta M_\mathrm{out, hot}$ & hot gas mass that enters outflow \\
    $\delta M_\mathrm{heat}$ & mass that transfers from cold to hot gas \\
    $\delta M_\mathrm{*}$ & stellar mass that forms per sub-step \\
    $\delta M_\mathrm{acc, hot}$ & hot gas mass that is accreted from IGM \\
    $\delta M_\mathrm{metals}$ & ejected metals by SNe per sub-step\\
    \\
    & parameters of the star formation model\\
    \hline
    $M_\mathrm{min}$, $M_\mathrm{max}$ & limiting masses of the IMF \\
    $\alpha$ & slope of the IMF \\
    $\eta$ & star formation efficiency \\
    $f_\mathrm{esc}$ & ionizing escape fraction \\
    $n^\mathrm{den}_\mathrm{cold}$ & number density of dense gas \\ 
    $v_\mathrm{BC}$ & baryonic streaming velocity \\
    $\gamma_\mathrm{out}$ & outflow efficiency (Eq.~\ref{eq:alphaout})\\
    $\alpha_\mathrm{out}$ & exponent in $\gamma_\mathrm{out}$\\
    $M_\mathrm{out,norm}$ & normalization mass in $\gamma_\mathrm{out}$ \\
    \\
    & times and timescales \\
    \hline
    $\Delta t_z$ & time between redshift steps \\
    $\delta t_i$ & adaptive timestep $i$ in the SF subcycle \\
    $t_\mathrm{dyn}$ & dynamical time scale of the halo \\
    $t_\mathrm{cold, ff}$ & free-fall time scale of cold gas \\
    $t_\mathrm{star}$ & stellar formation time scale \\
    $t_\mathrm{cool}$ & cooling time scale \\ 
    $t_\mathrm{acc}$ & hot gas accretion time scale \\
\end{tabular}
\caption{Definitions of the variables used in \asloth. The subscripts $_\mathrm{II}$ and $_\mathrm{III}$ in the main text refer to Pop~II- and Pop~III-specific quantities, respectively.}
\label{table:notation}
\end{table}

Each halo is initially assigned a baryonic mass fraction equivalent to the mean cosmic baryon fraction, i.e., the initial baryonic mass of a halo with virial mass $\mvir$ is assumed to be $(\Omega_\mathrm{ b} / \Omega_\mathrm{ m})\mvir$ prior to any star formation. While baryonic streaming might lower this baryon mass in low-mass halos at high redshift, this effect is not important for star-forming halos in our fiducial model \citep{naoz12}.
To model the evolution of this baryonic mass, we adopt a model in which the baryonic content associated with a halo is divided into four components: cold gas ($M_\mathrm{cold}$), hot gas ($M_\mathrm{hot}$), stars ($M_\mathrm{*}$), and outflows ($M_\mathrm{out}$). We summarize the nomenclature in Table~\ref{table:notation}.
During the evolution we add hot gas to a halo until
\begin{equation}
\begin{aligned}
    \mvir \frac{\Omega_\mathrm{b}}{\Omega_\mathrm{m}} = M_\mathrm{cold} + M_\mathrm{hot} + M_\mathrm{*} + M_\mathrm{out},
\end{aligned}
\end{equation}
which implies that halos accrete additional baryons whenever their DM mass increases by smooth accretion. Smooth accretion here refers to increases in DM mass and the associated increase in baryonic mass which is not caused by mergers or caused by mergers with unresolved halos. During mergers all baryonic mass components from the merging halos are added. If DM is accreted via smooth accretion, the baryonic mass budget of the halo is increased accordingly. The additional mass is gradually added to the hot component of the halo. Note that in our model, hot gas is any gas that is not cold, i.e.\ it corresponds to the sum of the warm and hot phases of the ISM in the usual three-phase description \citep{McKee77, Klessen16}. 
The outflow mass is the cumulative mass that is unbound from the halo and is never re-accreted. 

In addition, we allow star formation to have multiple epochs instead of one star burst in each global timestep $\Delta t_{z}$. To accurately follow star formation within individual halos we introduce sub-cycling. This is necessary, because the global timestep of most available merger trees is larger than the typical lifetime of massive stars, and so multiple epochs of stellar birth and feedback are expected within one global model timestep. At each global timestep, adopted from the time-evolution of the merger-trees, the halo inherits the baryonic properties from all of its progenitors, which are 
$M^{0}_\mathrm{cold}$, $M^{0}_\mathrm{hot}$, $M^{0}_\mathrm{*,II}$, and $M^{0}_\mathrm{out}$. 
Note that halos at the very first global timestep are initialized with hot gas only due to the accretion shock.
Then, in each sub timestep $i$, the baryonic content is updated using the following equations:
\begin{equation}
\begin{aligned}
    & M^{i+1}_\mathrm{ cold} =  M^{i}_\mathrm{ cold} + \frac{\delta t_{i} M^{i}_\mathrm{ hot}}{t_\mathrm{dyn}} - \delta M^{i}_\mathrm{ out, cold} - \delta M^{i}_\mathrm{ heat} - \delta M^{i}_{*}\\
    & M^{i+1}_\mathrm{ hot} = M^{i}_\mathrm{ hot} - \frac{\delta t_{i} M^{i}_\mathrm{ hot}}{t_\mathrm{ dyn}}  - \delta M^{i}_\mathrm{ out, hot} + \delta M^{i}_\mathrm{ heat} + \delta M^{i}_\mathrm{ acc, hot} \\
    &M^{i+1}_\mathrm{ out} = M^{i}_\mathrm{ out} + \delta M^{i}_\mathrm{ out, cold} + \delta M^{i}_\mathrm{ out, hot}, \\
\end{aligned}
\end{equation}
where we assume that the time scale of the cooling of hot gas is comparable to the dynamical time scale of the disk, which is composed of stars and cold gas in a central region with a radius of $R_\mathrm{s}$. Therefore, $t_\mathrm{dyn} = R_\mathrm{s} / v_\mathrm{dyn}$, $R_\mathrm{s} = R_\mathrm{vir} / c_\mathrm{dm}$ is the scale radius of a halo \citep{Navarro1996}, and the velocity
\begin{equation}
\label{eq:tdyn}
v_\mathrm{dyn} = \sqrt{G (M_{*}+M_\mathrm{cold}) / R_\mathrm{s}}
\end{equation}
is the characteristic velocity of the central part of the halo. For the halo's concentration, $c_\mathrm{dm}$, we use the fitting functions provided by \citet{Correa15}, which are provided in Appendix~\ref{app:cdm}. We describe how we determine each baryonic component in the following sections.

\subsection{Star Formation and Feedback}
\label{sec:sf}
\subsubsection{Critical Halo Mass for Cooling and Star Formation}
To form stars, gas needs to cool down efficiently. Since we cannot follow the chemical evolution of the gas without information on the spatial distribution of the baryons inside a halo, we instead use a simple mass criterion: rapid gas cooling and star formation is assumed to only occur in halos that have masses greater than a critical mass $M_\mathrm{crit}$. The appropriate value to take for $M_{\rm crit}$ is somewhat uncertain, with different studies in the literature yielding different results \citep{Yoshida03,latif19,Schauer21,kulkarni21}. \asloth\ therefore offers three different models for this critical mass: one based on \citet{Schauer21} (hereafter S21, our fiducial model); a second one based on \citet{Oshea08}, \citet{Hummel12} and \citet{Stacy11} (OHS); and a model based on \citet{Fialkov13} (F13). We describe the details of our fiducial model here. The other two models are described in Appendix~\ref{app:Mcrit}. 

In our fiducial model, we follow Eqs.~9 and 10 from \citet{Schauer21}, which quantify the dependence of $M_{\rm crit}$ on the strength of the LW background and the velocity associated with the large-scale streaming of the baryons relative to the DM, $v_\mathrm{ BC}$. Combining these equations yields the following expression for $M_{\rm crit}$:
    \begin{equation}
    \begin{aligned}
        &\mathrm{log_{10}} (M_\mathrm{ crit, S21}/\Msun) \\
        &= 6.0174 \, ( 1.0 + 0.166 \sqrt{J_{21}} ) + 0.4159 \, v_\mathrm{BC}\,.
    \end{aligned}
    \end{equation}
Here, $J_{21}$ is the strength of LW background in units of $10^{-21} \: \mathrm{erg \, s^{-1} \, cm^{-2} \, Hz^{-1} \, sr^{-1}}$ and $v_\mathrm{BC}$ is the streaming velocity in units of the root-mean-square value, $\sigma_{\rm rms}$. We adopt $v_\mathrm{BC}=0.8$, as this is the most likely strength of the streaming velocity \citep{Schauer21}, and we assume a redshift-dependent form for the LW background following \citet{GreifBromm06},
    \begin{equation}
        J_{21} = 10^{2-z/5}\,.
    \end{equation}

We assume a global LW background, because the effective volume that we simulate with \asloth\ is smaller than the mean free path of LW photons \citep{Ahn09}. In this public release, we do not include near-field LW feedback, which could self-regulate Pop~III star formation. However, users can easily implement this feature, for example following the model in \citet{hartwig16a}.

We note that the expression for $M_{\rm crit}$ we use here corresponds to the mass scale at which an average halo of that mass can cool and form stars. The minimum halo mass for cooling, $M_{\rm crit, min}$ is typically a factor of a few smaller \citep{Schauer21}, but as cooling and star formation only occurs in a few halos in the mass range $M_{\rm crit, min} < M < M_{\rm crit}$, we consider that $M_{\rm crit}$ is the more appropriate value for our purposes.
    
If a halo is above the atomic cooling limit ($T_\mathrm{ vir} \geq 10^4~$K), the gas can cool efficiently as long as the ionizing background is not strong \citep[][also see Section \ref{sec:r_ion}]{Visbal17}. We account for this via another critical mass scale
 \citep{Hummel12}
    \begin{equation}
        M_\mathrm{ACH}/\Msun = 10^{7.5} \left( \frac{1+z}{10} \right)^{-1.5}.
    \end{equation}
The final critical mass scale is then simply given by
\begin{equation}
M_{\rm crit, final} = {\rm min} \left(M_{\rm crit}, M_{\rm ACH} \right).
\end{equation}
    
\subsubsection{Stochastic Sampling of Stars}
Earlier versions of \asloth\ considered star formation only in clusters with IMF-averaged and continuous feedback \citep{Hartwig15b,Magg18,Tarumi20}.
We have improved the star formation model by adding a new implementation such that we keep track of individual stars \citep{chen22}. We store the numbers, masses, and birth times of stars that formed in each timestep. This novel feature of \asloth\ allows us to account for stellar feedback at specific times and to correctly model the stochastic feedback. In addition, we are also able to keep track of those stars which survive until $z=0$ and their stellar properties.

First, we determine whether we form Pop~III stars or Pop~II stars based on the empirically derived critical metallicity of \citet{Chiaki17}. We form Pop~II stars if
\begin{equation}
    10^{\mathrm{[C/H]}-2.30} + 10^{\mathrm{[Fe/H]}} > 10^{-5.07},
\end{equation}
where [Fe/H] and [C/H] are the iron and carbon abundances of the star-forming gas.
Then, we estimate the newly-formed stellar mass given a star formation efficiency $\eta_*$: 
\begin{equation}
    \label{eq:popiisf}
        M^{i}_\mathrm{ *, est} = \eta_* M^{i}_\mathrm{ cold} \frac{\delta t_{i}}{t^{i}_\mathrm{ cold, ff}}, 
\end{equation}
where $\eta_*$ is the star formation efficiency per free-fall time, $t^{i}_\mathrm{ cold, ff} = (G \rho^{i}_\mathrm{ cold})^{-1/2}$ is the free-fall time of cold gas, $\rho^{i}_\mathrm{ cold} = M^{i}_\mathrm{ cold} / V_\mathrm{ cold}$ is the mean cold gas density of the halo, $M^{i}_\mathrm{ cold}$ and $V_\mathrm{ cold}$ are the mass and volume of cold gas, respectively.

Next, we compute the average number of stars in the logarithmically spaced IMF bins. For Pop~II stars, we assume a Kroupa IMF and for Pop~III stars, we assume an IMF with slope $\alpha_\mathrm{III} = 1$. If $n_{j, \mathrm{avg}} \leq 10$, we determine the number of stars in bin $j$ with Poisson sampling. We deactivate Poisson sampling when $n_{j, \mathrm{avg}} > 10$ to save computational time.

\subsubsection{Photoheating}
Massive stars with $M_\mathrm{star} \geq 8\Msun$ dominate the production of energetic photons in a stellar population. These photons can ionize hydrogen, leading to the heating of gas. We compute a mass conversion rate for stars at different masses by estimating the mass that is enclosed in the HII region of a massive star. Our calculation assumes that the expansion of the HII region is governed by the \citet{Spitzer78} solution
\begin{equation}
\begin{aligned}
R_{\rm I}(t) = R_{\rm D} \left[ 1 + \frac{7}{4} \frac{c_{\rm s}(t-t_{\rm D})}{R_{\rm D}} \right]^{4/7}, 
\end{aligned}
\end{equation}
where $R_{\rm I}(t)$ is the radius of the region at time $t$, $R_{\rm D}$ is the radius at which the ionization front first undergoes the transition from R-type to D-type expansion, $t_{\rm D}$ is the time at which the front reaches $R_{\rm D}$ and $c_{\rm s}$ is the sound speed of the ionized gas. The latter is given by $c_\mathrm{s} = 11.4 \left( \mathrm{T}_\mathrm{ion}/{10^4\,\mathrm{K}} \right)^{1/2}$\,km/s, and we adopt a value  $\mathrm{T}_\mathrm{ion} = 10^4$\,K for the temperature of the ionized gas.\footnote{In reality, we expect gas in HII regions to have a range of different temperatures, depending on the metallicity of the gas and of the massive stars, the age and evolutionary state of the region, etc. However, in general these temperatures will differ from $10^{4}$~K by no more than a few tens of percent \citep[see e.g.][]{Ho2019}, and so the error we make by adopting the same fixed value for all HII regions will be small.}
Using the Spitzer solution, we derive\footnote{Details of this derivation can be found in Appendix B of \citet{chen22}.} the following expression for the instantaneous mass conversion rate:
\begin{equation}
    \dot{M}_\mathrm{heat} = m_\mathrm{H} n_\mathrm{cold,den} R_\mathrm{D}^2 c_\mathrm{s}\left[ 1 + \frac{7}{4} \frac{c_\mathrm{s}(t-t_\mathrm{D})}{R_\mathrm{D}} \right]^{-1/7}, 
    \label{eq:IoM_Main}
\end{equation}
where $m_\mathrm{H}$ is the mass of a hydrogen atom, and we assume $n_\mathrm{cold,den} = 10^3 \: {\rm cm^{-3}}$ for the number density of the neutral gas surrounding a newly-formed star. As star formation correlates with gas density \citep{GaoSolomon2004}, we expect this value to be larger than the mean density of the cold gas. Our choice of value here is motivated by \citet{Roman-Duval2010}, but in practice $\dot{M}_{\rm heat}$ is only weakly sensitive to this choice ($\dot{M}_\mathrm{heat} \propto n^\beta_\mathrm{cold,den}$ with $-3/7 \leq \beta \leq -1/3$). For other physical quantities related to cold gas mass such as the gas binding energy and the cold gas free-fall time, we use the mean cold gas density which is based on the current cold gas mass. 
We find that the mass conversion rate stays almost constant throughout the stellar lifetime (see Fig.~B1 in \citealt{chen22}). To save computational time, we therefore use a time-averaged mass conversion rate, $\langle \dot{M}_\mathrm{ heat}\rangle$. 

The radius of the ionization front at the time that it undergoes the transition from R-type to D-type is simply the initial value for the Str\"omgren radius: 
    \begin{equation}
    R_\mathrm{ D}= \left( \frac{3 Q}{4 \pi n_\mathrm{cold,den}^2 \alpha_\mathrm{ B}} \right)^{1/3},
    \end{equation}
where $\alpha_{\rm B}$ is the case B recombination coefficient, which for our adopted temperature of $10^{4}$~K is given by $\alpha_\mathrm{ B}=2.6\times10^{-13} \: {\rm cm^{3} \, s^{-1}}$  \citep{Ferland92}. The time required for the ionization front to reach $R_{\rm D}$ is given by
    \begin{equation}
    \begin{aligned}
    t_\mathrm{ D} = \ln\left[\frac{842}{23} \left(\frac{Q}{10^{48}\, \mathrm{s}^{-1}}\right)^{1/3} \left(\frac{n_\mathrm{ cold,den}}{10^3\,\mathrm{cm}^{-3}}\right)^{1/3}\right] \tau_\mathrm{recomb}, 
    \end{aligned}
    \end{equation}
    where $\tau_\mathrm{recomb} = (\alpha_\mathrm{ B}n_\mathrm{cold,den})^{-1}$ is the recombination time. 
    
    In dense molecular gas, stars form in clusters. Star clusters are less efficient in heating up the surrounding regions than isolated stars because their ionizing regions overlap. Therefore, we assume that 90\% of massive stars in a halo form in one big cluster at the galactic center, whereas the other 10\% of massive stars form in isolation \citep{chen22}. We sum up the ionizing photons from the 90\% of massive stars and compute one mass conversion rate $\dot{M}_\mathrm{ heat, cl}$ for the star cluster and compute individual $\langle \dot{M}_\mathrm{ heat, iso} \rangle$ for isolated stars. 
    Therefore, in timestep $i$, the total mass conversion rate from massive stars is given by 
    \begin{equation}
        \delta M_\mathrm{ heat, i} = \left( \dot{M}_\mathrm{ heat, cl} + \sum^{0.1N}_{j=1} \langle \dot{M}^{j}_\mathrm{ heat, iso}\rangle \right) \delta t_{i}, 
    \end{equation} 
    where $N$ is the total number of massive stars.
    For the full derivation and the two extreme cases, see the Appendix of \citet{chen22}.
    
    \subsubsection{Outflows Driven by Supernovae}
    We compute the gas mass ejected from the halo by supernovae by comparing the binding energy of gas with the total energy transferred to the gas by the supernovae. 
    DM is the dominant source of the gravitational potential, which determines the gas binding energy. The full three-dimensional DM distribution within the halos is not provided in the merger trees, and we therefore assume that all halos follow the NFW profile \citep{Navarro1996}.
    We compute the gas binding energy by considering contributions from DM, cold and hot gas, and stellar components. The gravitational potential of a halo is dominated by DM, however, we find that the contribution of baryonic content is non-negligible.
    We assume that the gas surrounding massive stars that die as supernovae has been heated up during their stellar lifetimes. Therefore, supernovae in \asloth\ eject the hot gas first. The binding energy of hot gas is computed with the formula:
    \begin{equation}
    \begin{aligned}
    & \mathrm{ E}^{i}_\mathrm{ bind, hot} = \frac{3 \mathrm{ G} \mpeak M_\mathrm{ hot,i}}{R_\mathrm{ vir} 
    \left[ \frac{-R_\mathrm{ vir}}{R_\mathrm{ s} + R_\mathrm{ vir}} + \mathrm{ ln}\frac{R_\mathrm{ s} + R_\mathrm{ vir}}{R_\mathrm{ s}} \right]} \times\\
    & \left[ -\frac{1}{4} +\frac{1}{2} \left( 1-\frac{R_\mathrm{ s}^2}{R_\mathrm{ vir}^2} \right) \mathrm{ ln}\frac{R_\mathrm{ s}+R_\mathrm{ vir}}{R_\mathrm{ s}} +\frac{1}{2}\frac{R_\mathrm{ s}}{R_\mathrm{ vir}} \right] \\
    & + \left( \frac{3R_\mathrm{ s}}{2R_\mathrm{ vir}}-\frac{13R_\mathrm{ s}^3}{30R_\mathrm{ vir}^3} \right) \frac{ \mathrm{ G} M_\mathrm{ disk,i} M_\mathrm{ hot,i}}{R_\mathrm{ s}} + \frac{3}{5} \frac{\mathrm{ G}M_\mathrm{ hot,i}^2}{R_\mathrm{ vir}}, 
    \end{aligned}    
    \end{equation}  
    where $\mpeak$ is the peak virial halo mass, $R_\mathrm{s}$ is the scale radius of the DM halo \citep{Navarro1996}, $R_\mathrm{vir}$ is the virial radius of the halo and $M^{i}_\mathrm{disk}$ is the combined mass of stars and cold gas since we assume they reside in the same central region in the halo \citep{chen22}. We use the peak virial halo mass until the current redshift. This definition is more robust against the fluctuating results of the halo finder.
    
    Similarly, the binding energy of cold gas can be described by 
    \begin{equation}
    \begin{aligned}
    & \mathrm{ E}^{i}_\mathrm{ bind, cold} = \frac{3 \mathrm{ G} \mpeak M_\mathrm{ cold,i}} {4 R_\mathrm{ s}
    \left[ \frac{-R_\mathrm{ vir}}{R_\mathrm{ s} + R_\mathrm{ vir}} + \mathrm{ ln}\frac{R_\mathrm{ s} + R_\mathrm{ vir}}{R_\mathrm{ s}} \right]}  \\
    & + \frac{6\mathrm{ G} M_\mathrm{ cold,i} \mste}{5R_\mathrm{ s}} + \frac{\mathrm{ G} M_\mathrm{ hot,i} M_\mathrm{ cold,i}}{R_\mathrm{ vir}} \left( \frac{3}{2} -\frac{3R_\mathrm{ s}^2}{10R_\mathrm{ vir}^2 } \right) \\
    & + \frac{3\mathrm{ G}M_\mathrm{ cold,i}^2}{5R_\mathrm{ s}}.
    \end{aligned}
    \end{equation}
    Complete derivations of both of these binding energies can be found in the Appendix of \citet{chen22}.

    We distribute the sum of SNe energies in this step, $E^{i}_\mathrm{SNe}$, to hot gas and cold gas based on their binding energies and masses. The fractions of $E^{i}_\mathrm{SNe}$ that affect hot gas and cold gas are, 
    \begin{equation}
    \begin{aligned}
        f_\mathrm{hot} & = \frac{\mathrm{E}^{i}_\mathrm{bind, hot}M^{i}_\mathrm{hot}}{\mathrm{E}^{i}_\mathrm{bind, hot}M^{i}_\mathrm{hot} + \mathrm{E}^{i}_\mathrm{bind, cold}M^{i}_\mathrm{cold}} \\
        f_\mathrm{cold} & = \frac{\mathrm{E}^{i}_\mathrm{bind, cold} M^{i}_\mathrm{cold}}{\mathrm{E}^{i}_\mathrm{bind, hot}M^{i}_\mathrm{hot} + \mathrm{E}^{i}_\mathrm{bind, cold}M^{i}_\mathrm{cold}}. \\       
    \end{aligned}    
    \end{equation}
    The amount of hot gas ejected by supernovae is calculated using the expression 
    \begin{equation}
    \label{eq:mouthot}
    \delta M^{i}_\mathrm{ out, hot} = \mathrm{ min} \left( \frac{ E^{i}_\mathrm{ SNe}f_\mathrm{hot}/\gamma_\mathrm{ out} }{E_\mathrm{ bind, hot}} M^{i}_\mathrm{ hot}, M^{i}_\mathrm{ hot} \right).
    \end{equation}
    
    Here, $\gamma_\mathrm{ out}$ is the outflow efficiency introduced by \citet{chen22} and defined as 
    \begin{equation}
    \label{eq:alphaout}
        \gamma_\mathrm{ out} =  \left( \frac{\mpeak}{M_\mathrm{out,norm}} \right)^{\alpha_\mathrm{out}},
    \end{equation}
    where the normalization mass $M_\mathrm{out,norm}$ and $\alpha_\mathrm{out}$ are free 
    parameters whose values are calibrated in Section~\ref{sec:calibration}.
    
    Similarly, the amount of cold gas ejected by supernovae is determined with  
    \begin{equation}
    \label{eq:moutcold}
    \delta M^{i}_\mathrm{ out, cold} = \mathrm{ min} \left( \frac{ E^{i}_\mathrm{SNe}f_\mathrm{cold}/\gamma_\mathrm{ out} }{E^{i}_\mathrm{ bind, cold}} M^{i}_\mathrm{ cold}, M^{i}_\mathrm{ cold} \right).
    \end{equation}

\subsection{Adaptive Timestep}
To evolve the baryonic physics in time, we numerically integrate the corresponding differential equations with an adaptive timestep. If chosen correctly, such an adaptive timestep saves computational time, while still allowing for accurate integration of dynamical effects that require a fine time resolution. The adaptive timestep is calculated for each halo at every time. The general idea is that we consider the main characteristic timescales of the system and then choose a timestep that is smaller than all these times. We consider the star formation timescale
\begin{equation}
    t_\mathrm{SF} = \frac{M_\mathrm{*}+10^2\Msun}{\dot{M}_\mathrm{*}},
\end{equation}
where $M_\mathrm{*}$ is the amount of stars already present in this halo, $\dot{M}_\mathrm{*}$ is the current SFR, and we add a small offset mass to prevent this timestep from getting too small, i.e., to gain computational speed.
These small offsets here and below are chosen empirically as the largest values that still guarantee a stable time integration under all conditions. We expect roughly one SN per $100\Msun$. So this offset corresponds to the smallest unit of stellar mass whose formation we want to resolve in time. A smaller timestep would mean that we follow the formation of SN progenitor stars with unnecessarily high time resolution.

The accretion time scale is given by
\begin{equation}
    t_\mathrm{acc} = \frac{M_\mathrm{hot}+10^4\Msun}{\dot{M}_\mathrm{acc}},
\end{equation}
where $M_\mathrm{hot}$ is the amount of hot gas already in the halo, $\dot{M}_\mathrm{acc}$ is the accretion rate of new hot gas from the IGM, and we add a small offset mass to prevent this timestep from getting too small. Here, the offset mass is chosen as $\sim 1\%$ of the baryonic mass of a typical minihalo. This means that we allow \asloth\ to approximate the accretion of hot gas at the percent level in order to gain computational efficiency.

We also use the dynamical time as defined in Eq.~\ref{eq:tdyn}. We found that the time integration of cold gas can be improved when we use $0.1t_\mathrm{dyn}$ as a constraint for the dynamical timestep.

The merger tree is defined based on redshift steps, which is the timescale on which halos merge. To account for this, we also include the time between redshift steps, $\Delta t_z$, in our calculation of the adaptive timestep.

The adaptive timestep is then given by
\begin{equation}
    t_\mathrm{adapt} = 0.025 \min(t_\mathrm{SF},t_\mathrm{acc},0.1t_\mathrm{dyn},\Delta t_z) + 1\,\mathrm{yr}.
\end{equation}
The addition of one year prevents extremely small timesteps in some pathological cases and the prefactor of $2.5\%$ is the largest value that still guarantees a stable and converged evolution of the baryon budget.

The smallest of these timescales, which then defines the adaptive timestep depends on redshift, halo mass, and other factors. \asloth\ provides routines to visualize the relative importance of these characteristic timescales in a specific setup.

\begin{figure*}
    \centering
    \includegraphics[width=\textwidth]{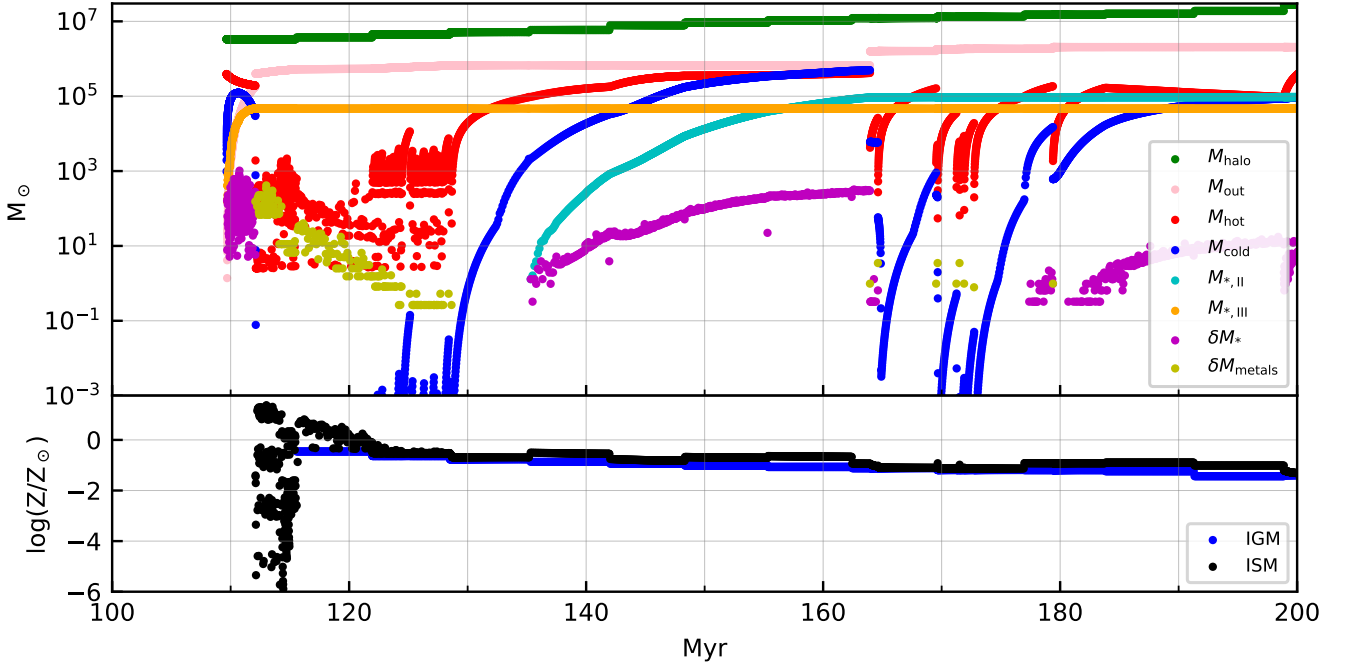} 
    \caption{Illustration of the baryonic sub-cycling and what physical quantities are available within \asloth. Top: various mass contents along one merger tree branch as a function of time after the Big Bang. Bottom: gas metallicity of the IGM (blue) and ISM (black). The illustrated time corresponds to the redshift range $z=27-18$. We witness the formation of the first Pop~III stars around 110\,Myr (orange), the suppression of further star formation by SNe around 120\,Myr (olive), the accumulation of further cold gas around 130\,Myr (blue), and the formation of second-generation stars after $\sim 140$\,Myr (cyan). The ISM metallicity experiences strong variations during the first SN explosions. This is due to the small amount of residual gas and metals that remain in the halo. The new stellar mass $\delta M_*$ (purple) and the ejected metal mass $\delta M_\mathrm{metals}$ (olive) are the instantaneous values at the specific timestep during subcycling. The halo mass $M_\mathrm{halo}$ (green) increases in steps, which illustrates the original redshift steps from the merger tree. The stellar masses (orange, cyan) and outflow mass (pink) are cumulative and therefore monotonically increasing.}
    \label{fig:tgas}
\end{figure*}

Fig.~\ref{fig:tgas} shows how the different gas phases of our model evolve over time with an adaptive timestep.
This figure is an example for the main branch of a merger tree leading to the formation of a MW-like galaxy and is therefore not representative of the behavior of all minihalos. It illustrates how different baryonic components are converted into each other in a minihalo in the redshift range $z=27-18$.

\subsection{Spatial Feedback}
\label{sec:SpatialFeedback}
One of the primary advantages of using merger trees from $N$-body simulations is that the position of halos are known and that sub-halos are resolved. This opens up the possibility of modeling the spatially resolved reionization and enrichment history of the IGM. The method for this was introduced in \citet{Magg18} and the algorithm was refined in \citet{Magg21b}. For completeness, we describe the implementation in this Section. We provide further technical details on the implementation of spatial feedback in Appendix~\ref{app:bubble}.

\subsubsection{External Enrichment and IGM Metallicity}
Each star-forming halo $i$ is assigned an enriched region of radius $R_{\mathrm{en},i}$. Outflows are launched at the virial radius, i.e., $R_{\mathrm{en},i}$ is initialized to be the virial radius. Here, outflows refer to baryons that are pushed outside the virial radius of a halo.

Following \citet{agarw12}, we assume an outflow velocity of $v_\mathrm{out} = 110\,\mathrm{km}\,\mathrm{s}^{-1}$. We chose this fixed value, as due to the calibrated efficiency factor in Eq.~\ref{eq:alphaout}, we cannot self-consistently compute the outflow velocity from momentum conservation or energetic consideration. This efficiency factor was required to be able to match the stellar mass to halo mass relation. The mass of these outflows is modeled according to Eqs~\ref{eq:mouthot} and \ref{eq:moutcold}. The expansion of the outflow is modeled as a pressure-driven snowplow. In this approximation the enriched region expands with constant momentum and slows down as it sweeps up the IGM. The expansion velocity is
\begin{equation}
 v_\mathrm{en} = v_\mathrm{out}\frac{M_\mathrm{out}}{M_\mathrm{out}+\frac{4}{3}\pi \rho_\mathrm{b}\left(R_\mathrm{en}^3-\Rv^3\right)}.
\end{equation}
where $\rho_\mathrm{b}$ is the mean baryon density of the IGM, i.e.,
\begin{equation}
 \rho_\mathrm{b} = \Omega_\mathrm{b} \frac{3 H_0^2}{8\pi G}(1+z)^3.
\end{equation}
The enriched regions also increase in volume due to cosmic expansion. When two halos with enriched bubbles around them merge, we add their volumes and their matter  and metal contents. 

A halo $k$ is subject to external enrichment if its center is within the enriched bubble of any halo. We compute the mean IGM metallicity at the position $\vec{x}$ of a halo as
\begin{equation}
\label{eq:ZIGM}
\begin{split}
    &Z_{\mathrm{IGM},k} =\\ &\frac{\sum_{j \in \mathcal{E}} \rho_{\mathrm{met}, \mathrm{out}, j}}{\rho_\mathrm{b} + \left< 3 \Omega_b \Omega_m^{-1} M_{\mathrm{vir},k}(4 \pi R_{\mathrm{en},j}^3)^{-1}\right>_{j \in \mathcal{E}} + \sum_{j \in \mathcal{E}} \rho_{\mathrm{out}, j}}
\end{split}
\end{equation}
where $\mathcal{E}$ is the set of enriching halos, i.e., the set of halos $j$ with a distance to $\vec{x}$ of less than $R_{\mathrm{en}, j}$. For each enriching bubble the mass density is
\begin{equation}
\rho_{\mathrm{out}, j} =3 M_{\mathrm{out}, j}(4 \pi R_{\mathrm{en},j}^3)^{-1},
\end{equation}
and the metal density is 
\begin{equation}
    \rho_{\mathrm{met}, \mathrm{out}, j}= \rho_{\mathrm{out}, j} Z_{\mathrm{out}, j}.
\end{equation}
The term $\left< 3 \Omega_b \Omega_m^{-1} M_{\mathrm{vir},k}(4 \pi R_{\mathrm{en},j}^3)^{-1}\right>_{j \in \mathcal{E}}$ in Eq.~\ref{eq:ZIGM} takes into account that the enriching bubbles have to encompass the halo that is being enriched. As we base the accretion of metals on the metallicity of the IGM, it is not limited by the mass of metals contained within the bubbles. Thus, without this term, a very small but high-metallicity bubble could lead to the accretion of more metals than it contains. With this correction term, the accretion shows the correct behavior: in the limit that the accretion is dominated by very massive bubbles and in the limit that the mass budget is dominated by the accreting halo, the accreted metal mass is limited to the available amount of metals. This method of correcting is necessary, as our algorithm currently does not allow to remove metals from the bubbles.

\subsubsection{Photoionization of the IGM}
\label{sec:r_ion}
In addition to enriching the IGM with metals, massive stars will also photoionize the surrounding IGM. We assign each halo an ionized region with radius $R_\mathrm{ion}$. Again, this radius is initialized to be the virial radius. For convenience, we phrase the expansion of each ionized region in terms of its volume $V$ and afterwards compute the corresponding radius as $R_\mathrm{ion} = \left(3V /4\pi \right)^{1/3}$. The change in ionized volume can be computed by modeling the expansion as an R-type ionization front, i.e., modeling the expansion assuming that the difference between the number of recombinations and the number of produced ionizing photons lead to expansion or shrinking of the ionized region. If we assume the gas inside the ionized region is fully singly ionized we find, 
\begin{equation}
    \dot{N}_\mathrm{ion} = V n^2 C\alpha_B + \dot{V} n
\end{equation}
for a nucleon number density of $n$ and an ionizing photon emission rate $\dot{N}_\mathrm{ion}$. Helium reionization is neglected in this model. We assume a clumping factor of $C=3$ \citep{Robertson13}. To ensure numerical stability, we integrate this equation with an implicit Euler method, and therefore we update the volume of ionized regions during a timestep $\Delta t _z$ from $k$ to $k+1$ as 
\begin{equation}
  V_{k+1} = \left(V_k +\frac{ \dot{N}_\mathrm{ion}\Delta t}{n}\right)\left(1+\Delta t_z n\alpha_B C\right)^{-1}.
  \label{Eq:Vion}
\end{equation}
Since the temperature of ionized gas in the IGM is around $T\approx 10^4\,\mathrm{K}$, halos with a virial temperature lower than this cannot gravitationally bind the ionized gas. Thus halos below the atomic cooling limit (i.e., $T_\mathrm{vir} = 10^4\,\mathrm{K}$) are not allowed to accrete gas from the IGM. We extend this to masses up to a factor of 10 above the atomic-cooling limit \citep{Visbal17} for halos that are exposed to large fluxes of ionizing photons, $F_\mathrm{ion} > 6.7\times10^5\,\mathrm{photons}\,\mathrm{s}^{-1}\,\mathrm{cm}^{-2}$.
If the center of one bubble lies within another they are joined in a new bubble, where the center of the new bubble is the center of mass of the previous two bubbles and the volumes are added. This method prevents a cluster of strong ionizing sources from ionizing the same volume multiple times, and instead creates one big joined ionized region. Similar to the enriched volumes, when two halos with ionized bubbles merge, we add the volume and increase it according to cosmic expansion.

We do not use a global, external ionizing background, but calculate the reionization self-consistently within our merger trees. Hence, we implicitly assume that the ionizing radiation feedback in our simulations is dominated by the local sources. The mean free path of ionizing photons in a neutral IGM at the redshift of reionization ($z\sim 7$) is only 3 comoving kpc, so this is a good approximation at early times, while the size of most ionized bubbles remains small. It breaks down once connected ionized regions with sizes comparable to the size of our simulation volume start to become common, as at this point there is a substantial likelihood of such a region expanding into our volume from the outside. Hydrodynamical simulations of reionization in large volumes \citep[e.g.][]{gnedin14,Rahmati18,kannan22} suggest that this happens once the ionized gas volume filling fraction exceeds $\sim 50$\%, and so it is likely that \asloth\ slightly underestimates the time taken to transition from this ionized fraction to a fully ionized state. In order to do a better job of modelling this late stage of reionization, it would be necessary to use merger trees drawn from a simulation of a much larger volume -- for instance, \citet{Iliev14} argue that a box size of at least $100 h^{-1} \: {\rm Mpc}$ is needed to obtain a fully converged reionization history. Unfortunately, we are not aware of a publicly available $N$-body simulation that is larger than the one that we use from \citet{Ishiyama16} but that also has the mass resolution and redshift coverage that we require for \asloth.  

Finally, we note that we do not yet include the contribution from AGNs to the reionization model. This should be a good approximation as most studies find their contribution to the ionizing budget prior to reionization to be negligible \citep[see e.g.][]{trebitsch21}.

\subsubsection{Metallicities and Mixing}
\label{sec:mixing}
Stars synthesize metals and contribute to chemical enrichment. We assume progenitor-mass specific SN yields for Pop~II and Pop~III stars separately, and calculate the mass of each element by multiplying the yields with the number of exploding SN. 
The synthesized metals can escape from the halo together with the outflow bubbles. We assume that the fraction of metals ejected from the galaxy is
\begin{equation}
    f_\mathrm{eject} = \frac{\delta M_\mathrm{out}}{M_\mathrm{hot}+M_\mathrm{cold}}.
\end{equation}
 This implementation might underestimate the metallicity of the outflow and hence of the IGM, because SN-driven outflows are usually more metal-rich than the average gas in the galaxy \citep{MacLow99,Emerick2018,pandya21}. However, our simple approximation is able to reproduce the relative number of EMP stars observed in the MW, as we show below.

It is also possible that the metallicity of star-forming gas in a galaxy could be significantly different from the mean gas-phase metallicity of the galaxy \citep{Xu16c, Magg21a}. \cite{Tarumi20} investigate this scenario by analyzing 3D hydrodynamic simulations and find that star-forming gas in halos that have not experienced star formation (purely externally enriched halos) has significantly lower metallicity than the mean gas metallicity. Metals that are accreted onto the halo can enrich the gas at the outskirts, but do not enrich the star-forming gas close to the center of the halo. To account for this, in \asloth, we define a metallicity shift factor $dZ$ as
\begin{equation}
     dZ=Z_\mathrm{sf}-Z_\mathrm{ave},
\end{equation}
where $Z_\mathrm{sf}$ is the metallicity of the star-forming gas and $Z_\mathrm{ave}$ is the average metallicity of the halo. The distribution of $dZ$ is different between internal enrichment and external enrichment \citep{Tarumi20}. For computational efficiency, we define a metal-enriched halo as externally enriched if the total metal mass from SNe inside this halo is $<10^{-15}\Msun$ (effectively zero). For internal enrichment, inhomogeneity is not significant. The distribution of $dZ$ in this case is a Gaussian function with ($\mu, \sigma$) = ($-0.03, 0.15$). In the case of external enrichment, $dZ$ is typically significantly negative. We use a fitting function that we have derived in our previous study \citep{Tarumi20} and that is provided in Appendix~\ref{sec:dZfit}. We apply this metallicity correction to the metallicity of the newly formed stars.

\begin{figure}
\includegraphics[width=\columnwidth]{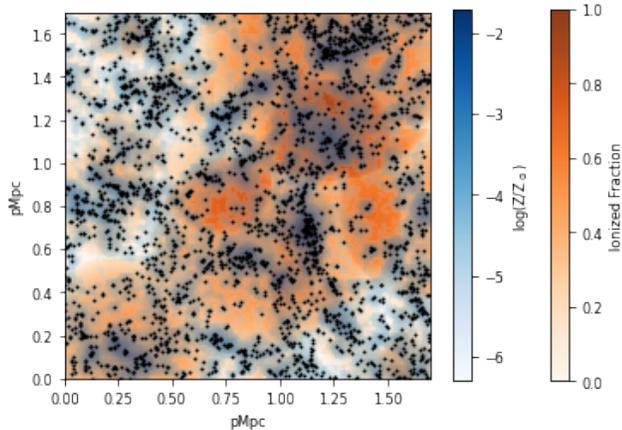}
\caption{\label{fig:projection}Illustration of the spatial extent of external feedback at $z=6$. We show the fraction of ionized volume along lines of sight in orange, the average IGM metallicity along lines of sight in blue, and halos with $\mvir>4 \times 10^8\Msun$ as black dots.}
\end{figure}

To illustrate the spatial feedback, we show a projection of the 8\,Mpc box in Fig.~\ref{fig:projection}.
This figure visualizes the spatial extent of radiative and chemical feedback in our cosmologically representative 8\,Mpc box. While this figure is a qualitative visualization to demonstrate the available information inside \asloth, we quantify the evolution of volume filling fractions over time below.

\subsection{Stochastic Feedback}
In some situations, it is not possible or desirable to follow the detailed spatial feedback. Either one wants to save memory and runtime, or one is using EPS-generated merger trees that do not have the information on the spatial positions of the halos that our spatial feedback implementation requires. For such cases, we have also implemented and tested an option for stochastic feedback that does not rely on spatial information but that still provides reasonable results.

The main idea is that we follow the volume filling fractions of metal enrichment and of ionizing radiation and then determine stochastically if a halo is currently in an ionized or metal-enriched region. Once a halo is in such a region, we assume that all its children are so as well. One could construct rare pathological cases in which this assumption does not hold, but we have found that this simplifying assumption adequately reproduces the desired behavior. 

The bubbles around each halo of the ionized and metal-enriched volume expand as described above. At each check, the probability that a halo is affected by external feedback is equal to the respective volume filling fraction at that time. Therefore, we only check for external feedback once, when the halo first surpasses $M_{\rm crit}$. Otherwise, if we would check for external feedback every timestep and make the timesteps arbitrarily short, the effective probability for external feedback would artificially increase because we would check more often. We show below that checking for external enrichment once provides the correct results. Moreover, we do not add metal masses from external metal enrichment in the case of stochastic feedback; we only flag the halo as externally enriched in such a case.

For any application for which metal enrichment and radiative feedback are important, we recommend users of \asloth\ to use spatially resolved merger trees from $N$-Body simulations instead of EPS-generated merger trees. We still provide this mode of \asloth, because it is a very efficient way to explore trends with input parameters, and it allows new users to immediately run simulations, without the need to download several GB of merger tree data.

\section{Calibration}
\label{sec:calibration}
In this section, we first introduce the six observables, based on which we calibrate \asloth. Then, we discuss our calibration strategy and how we found an optimal fit for the input parameters. We show how this calibrated model reproduces other constraints that \asloth\ was not incentivized to fit. Finally, we discuss how the model performs with EPS-generated merger trees.

\subsection{Six Observables}
We use six different observables to calibrate the input parameters of \asloth: the optical depth to Thomson scattering, the stellar mass of the MW, the cumulative distribution of stellar masses of the MW satellite galaxies, the fraction of EMP stars in the MW halo, the ratio of ultra metal-poor (UMP, [Fe/H] $<-4$) to EMP ([Fe/H] $<-3$) stars, and the cosmic SFRd at high redshift. For each of these observables, we calculate a goodness-of-fit parameter, $p_i$, which quantifies how consistent the output of \asloth\ is with observations. These parameters can have values in the range $0 \leq p_i \leq 1$, and higher values represent better fits. Some of these values are inspired by p-values (see below), for which a value of $p_i > 0.05$ conventionally means that a model is accepted (i.e. can not be rejected).

Eventually, we multiply these six individual parameters to obtain our final goodness-of-fit measure, $p_\mathrm{all}$. Maximizing $p_\mathrm{all}$ during the calibration guarantees that all six observables are reproduced simultaneously in the final model. In App.~\ref{sec:AppPara}, we present how the six observables depend on the input parameters. In Table~\ref{tab:fitted} we present the best-fit values, which will be discussed in more detail below.

\begin{deluxetable*}{lllr}
\tablecaption{Six observables based on which we calibrate \asloth, together with their observed values (column 2), modeled values (column 3), and their resulting goodness-of-fit parameters (last column). \label{tab:fitted}}
\tablewidth{0pt}
\tablehead{
\colhead{Observable} & \colhead{observed} & \colhead{\asloth} & \colhead{goodness-of-fit parameter} 
}
\decimalcolnumbers
\startdata
Optical Depth & $0.0544 \pm 0.0073$ & $0.0539$ & $0.989$  \\
$M _\mathrm{*,MW}$ & $(5.43 \pm 0.57) \times 10^{10}\Msun$ & $(7.3 \pm 3.0) \times 10^{10}\Msun$ & $0.529$  \\
CSMF & -- & -- & $0.136$  \\
log$_{10}$(EMP/All) & $-4.7\pm0.5$ & 
$-4.4 \pm 0.6$
& $0.697$  \\
log$_{10}$(EMP/UMP) & $(1.4 - 2)\pm0.3$ & 
$1.2 \pm 0.2$
& $0.563$  \\
SFRd & -- & -- & $0.479$  \\
 &  &  & $p_\mathrm{all}=0.0134$  \\
\enddata
\tablecomments{The statistical uncertainty for \asloth-based values is the standard deviation from 30 independent \ctp trees.}
\end{deluxetable*}

\subsubsection{Ionization History}
A key observable from the early Universe is the Thomson scattering optical depth $\te$. Free electrons can scatter CMB photons, leading to a slight polarization of the CMB. As only free electrons take part in this process, \te\ is influenced by the ionization fraction of the IGM as a function of time, i.e., the ionization history of the Universe. The \te\ parameter is computed as
\begin{equation}
 \te(z) = c \sigma_{\rm T} n_{\rm H} \int _0 ^z \mathrm{d}z' f_e Q_{\rm ion} (z') (1+z')^3 \left| \frac{\mathrm{d}t}{\mathrm{d}z'} \right|,
\end{equation}
where $z$ is the redshift, $\sigma_{\rm T} = 0.665 \times 10^{-24} \, \mathrm{cm}^2$ is the Thomson scattering cross-section, $n_\mathrm{H}$ the cosmological mean density of hydrogen nuclei at $z=0$, $Q_{\rm ion}$ the volume filling fraction of ionized regions, and $f_e$ the number of free electrons per hydrogen nucleus in the ionized IGM. The presence of helium makes this number slightly larger than one and we assume for simplicity that
\begin{equation}
 f_e = \begin{cases}
 1+Y_p/2X_p & \mathrm{at}\ z \leq 4 \\
 1+Y_p/4X_p & \mathrm{at}\ z > 4,
 \end{cases}
\end{equation}
where $Y_p$ and $X_p$ are the primordial abundances of He and H, respectively \citep{Robertson13}. This prescription assumes that He$^{+}$ is produced at the same time as H$^{+}$, primarily via radiation from stellar sources, and that the helium is later ionized further to He$^{++}$, at $z = 4$, as a consequence of the steadily increasing contribution of high-redshift AGN to the extragalactic UV background. In practice, $\te$ is only weakly sensitive to the redshift at which this occurs. The time evolution of the volume filling fraction is based on Eq.~\ref{Eq:Vion}.
\citet{Planck2016} determined\footnote{We adopt the value of $\te$ from \citet{Planck2016} for consistency with earlier work with \asloth, but note that the difference between this value and the most recently published value in \citet{Planck2020} is negligible.} the value of the Thomson scattering optical depth to be $\te=0.0544\pm 0.0073$. Notably, this value is much lower than what was reported earlier \citep[e.g.][]{wmap7year}, which implies much less star formation at the highest redshifts \citep{Visbal2015}. The instantaneous ionization redshift corresponding to this optical depth is $z=8.3$. We can compute \te\ from the ionization fraction of the Universe that we determine based on the total volumes of the ionized regions, as described in Section \ref{sec:r_ion}.

\begin{figure}
\includegraphics[width=\linewidth]{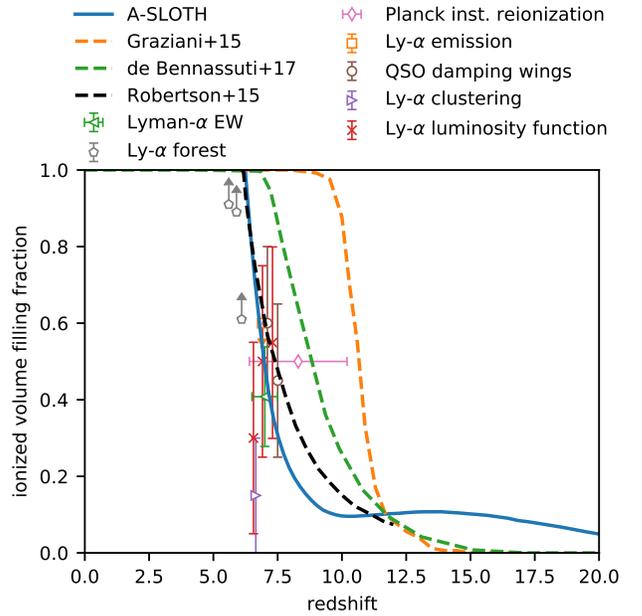}
\caption{\label{fig:tau}Ionization history simulated with \asloth\ (blue), compared to various models (dashed) and observations (points).}
\end{figure}

In Fig. \ref{fig:tau}, we show the ionization history derived from our 8\,Mpc box as compared to various observations and a selection of representative models from the literature \citep{robertson15,graziani15,deBennassuti17}.
The observational constraints are based on various different techniques \citep{McGreer15,mesinger15,Planck2016,greig17,zheng17,ouchi18,mason18,banados18,konno18}. The models by \citet{graziani15} and \citet{deBennassuti17} are derived for a MW-like volume, and are hence not directly comparable to our cosmologically representative box. We calibrate \asloth\ only based on the optical depth to Thomson scattering, i.e., the integrated reionization history. Our ionization history (blue line) agrees well with observational constraints on the volume filling factor of ionized gas at high redshift (the data points in the figure), which confirms the consistency of \asloth.

\subsubsection{Stellar Mass of MW and Its Satellites}
We calibrate \asloth\ to reproduce the observed MW stellar mass, and the number counts of satellites at $z=0$. To do this, we use the merger trees from the \ctp project simulations, each of which was designed to follow the build-up of a MW-like galaxy. The mean stellar mass we obtain with our fiducial \asloth\ model for these MW analogs is $\langle {M_\mathrm{star, MW}} \rangle = (7.3 \pm 3.0) \times10^{10}\Msun$. For comparison, the observed stellar mass of the real MW is $M^\mathrm{obs}_\mathrm{star, MW} = 5.43\pm0.57 \times 10^{10}\Msun$ \citep{McMillan2017}. 

\begin{figure}[ht]
\includegraphics[width=\columnwidth]{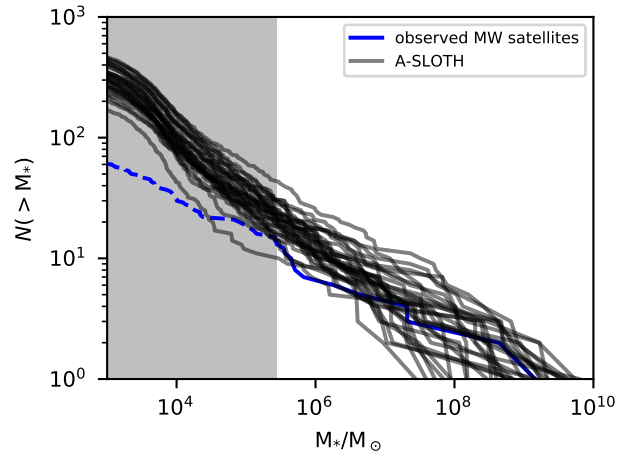}
\caption{CSMFs at $z=0$ extracted from our set of 30 \ctp trees, processed using our fiducial model parameters. The CSMFs from our \ctp trees are plotted in gray and the observed CSMF for the MW is plotted in blue. The gray shaded area represents the region in which the observations are known to be incomplete. The observed CSMF in this region is plotted using a dashed line and is not used to calibrate the model.  \label{fig:MstarCum}}
\end{figure}

In Fig.~\ref{fig:MstarCum}, we show the cumulative stellar mass function (CSMF) of satellites from the 30 \ctp trees, computed using our fiducial model. Each gray curve represents the CSMF from one tree, and for comparison the blue curve shows the observed CSMF from \citet{McConnachie2012} and \citet{Munoz2018}. Due to the limit of current surveys, the search for MW satellites is incomplete below $M_* \sim 10^5 \Msun$. We shade the region below this mass in gray and only calibrate our results to the observed CSMF above this observational completeness limit. We note that these compilations of satellite galaxies do not yet include AntII and CraII, which both have stellar masses of $\sim 10^6\Msun$ \citep{ji21}, but including these two galaxies would not significantly change our results.

To quantify the difference between our predicted CSMFs and the observed one, we use a two-sample Kolmogorov-Smirnov test. We perform the tests on each CSMF produced by \asloth\ and the observed CSMF. From each test, we retrieve a p-value, which we use as a goodness-of-fit parameter. We then compute the average of these parameters from the 30 tests.

 We note that, by construction, we neglect the effects of tidal stripping. While this effect is important for at least some of the nearby satellites \citep{Simon2018}, there are several reasons why it cannot be accurately modeled in our current framework. Firstly, both the shape of the potential well of the MW and of the satellites are inaccurate without accounting for the baryonic component and without the presence of the MW disk. Thus the tidal stripping cannot be accurately modeled in the dark-matter-only simulations that we base our model on \citet{Bullock2017}. Secondly, halo finders often struggle to correctly identify subhalos and their masses. Indeed, in many cases we see a substantial increase in the mass of the subhalo during tidal stripping events, which is caused by the halo finders associating some of the mass of the main halo to the subhalo. Furthermore, the effect of tidal stripping is more significant on satellites that are in proximity to the MW. Thus we neglect tidal stripping in the current work and leave it to future versions, ideally built on merger trees with disk potentials to improve on this aspect.

\subsubsection{Extremely Metal-Poor Stars}
We use two metallicity-related values for the calibration. These are the fraction of stars in the MW that are EMP stars, $f_\mathrm{EMP}$, and the ratio between the fraction of UMP stars and the fraction of EMP stars, $f_\mathrm{UMP}/f_\mathrm{EMP}$. We focus on these two individual values rather than comparing with the full MDF because it is easier to combine the values obtained from several different surveys (see below) and because it gives us a straightforward way of quantifying our agreement with the observations. We assume that EMP stars in \asloth\ are predominantly found in the halo \citep{kobayashi11}. Also, \citet{arentsen20} did not find any EMP stars among 8000 spectroscopically observed bulge stars. However, \citet{Sestito2019} found several EMP stars with disk-like orbits.

For the EMP fraction, we assume that (i) the stellar mass in the MW is $6\times 10^{10} \Msun$ \citep{McMillan2017}; (ii) the stellar mass in the halo of the MW is $1\times 10^{9} \Msun$ \citep{BullockJohnston05}; and (iii) 1 out of 800 halo stars is an EMP star \citep{Youakim20}. For the UMP to EMP ratio, we assume three possible values based on the literature. \cite{Youakim20} find in the Pristine survey that $10^{-2}$ of EMP stars are UMP stars. \cite{chiti21} argue that the MDF $dN/dZ$ can be well-fitted with a power-law function $Z^{-1.5}$, implying that $f_\mathrm{UMP}/f_\mathrm{EMP} = 10^{-1.5}$. Finally, \cite{bonifacio21} use Gaia and SDSS photometry in the TOPoS survey to derive a UMP to EMP ratio of $8/217 \simeq 10^{-1.4}$.

To evaluate the goodness-of-fit, we calculate how significantly the two metrics ($f_\mathrm{EMP}$ and $f_\mathrm{UMP}/f_\mathrm{EMP}$) are away from these observations. We outline here the procedure we use in the case of $f_\mathrm{EMP}$ but note that the treatment is similar for $f_\mathrm{UMP}/f_\mathrm{EMP}$. For the 30 \ctp merger trees, we first calculate the mean value of the logarithm of $f_{\rm EMP}$, $\mu_\mathrm{CTP} = \mathrm{mean}[\log_{10}(f_\mathrm{EMP})]$, and its standard deviation $\sigma_\mathrm{CTP} = \mathrm{stddev}[\log_{10}(f_\mathrm{EMP})]$. Then we calculate the ``error'' from an observation $i$ as $f_{i} = |\mu_\mathrm{CTP}-\log_\mathrm{10}[f_\mathrm{EMP, i}]|/\sigma$. Here $i$ stands for the three observations that we compare to, and $\sigma = \sqrt{(0.5)^{2}+\sigma_\mathrm{CTP}^{2}}$ is the error estimate that includes 0.5 dex observational uncertainty. The main source of this large uncertainty is the stellar mass estimate in MW halo that ranges from $4\times 10^{8} \Msun$ \citep{Bell2008} to $10^{9} \Msun$ \citep{BullockJohnston05}. Then we use $\mathrm{min}(f_\mathrm{Pristine}, f_\mathrm{Chiti}, f_\mathrm{TOPoS})$ as the goodness-of-fit evaluator.

Our approach predicts a mean EMP fraction of $\log_{10}[f_\mathrm{EMP}] = -4.39$ together with a mean EMP to UMP fraction of $\log_{10}[f_\mathrm{EMP}/f_\mathrm{UMP}] = -1.23$. The value of $\log_{10}(f_\mathrm{EMP})$ that we obtain lies $0.40 \sigma$ away from the observed value. The value of $\log_{10}(f_\mathrm{UMP}/f_\mathrm{EMP})$ lies $0.58 \sigma$ away from the \cite{bonifacio21} value and around one sigma away from the \cite{chiti21} value. We consider these to be good fits, which provide the second and third highest goodness-of-fit parameters out of the six observables.

\subsubsection{Cosmic Star Formation Rate Density}
The cosmic SFRd quantifies how many stars form per unit volume per unit time in the Universe. \asloth\ can predict the SFRd at high redshifts and surveys can estimate the SFRd out to $z \sim 10$. Since the SFRd is a cosmic average, we use the 8\,Mpc box as a cosmologically representative sample of the Universe. We only have DM halo merger trees for the 8\,Mpc box down to redshift $z \sim 5$. So for this observable, we compare the SFRd from \asloth\ to observations in the redshift range $5 \leq z \leq 10$.

The SFRd cannot be observed directly. Instead, galaxy counts at high redshift need to be de-biased and the galaxy mass function needs to be extrapolated to the faint end at different redshifts. Hence, the observation-based SFRd is not well constrained at high redshift. Therefore, we will compare our model with two extreme cases. Specifically, we compare our modeled SFRd to the predictions by \citet{Madau14} and \citet{behroozi19}. These two estimates differ by up to 1\,dex at high redshift. We require that \asloth\ lies between these observations or penalize the final goodness-of-fit parameter if our prediction is above both or below both observations.

We compare the maximum difference between our model and observations by using the cumulative stellar mass density that has formed up to a certain redshift. To simplify the interpretation, we calculate the cumulative stellar mass density as function of cosmic time (and not redshift), so that the resulting unit is stellar mass per unit volume. We compare these cumulative stellar mass densities in the range 300\,Myr - 1300\,Myr after the Big Bang, which is given by the highest redshift for which we have observational constraints and the lowest redshift of the DM merger tree.

\begin{figure}[ht]
\includegraphics[width=\columnwidth]{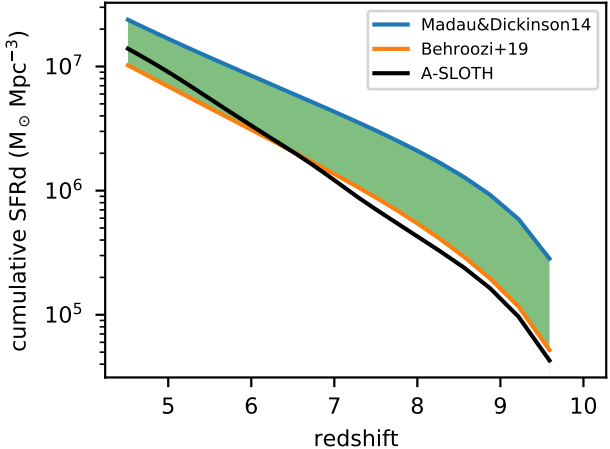}
\caption{Best \asloth\ model (black) compared to two independent observation-based constraints. We incentivize the result to lie between these two constraints and penalize the maximum distant that our modeled cumulative SFRd is outside the green band. \label{fig:cumSFR}}
\end{figure}

Fig.~\ref{fig:cumSFR} shows the two observation-based constraints and our modeled prediction.
At higher redshift, the SFRd of \asloth\ is below the prediction of both \citet{Madau14} and \citet{behroozi19}. At each time, we measure how far we are outside the band that is allowed by observations and we normalize it by the width of this band. Then, we quantify the discrepancy by choosing the time at which this relative discrepancy is largest. In this case, we would then calculate the relative distance as
\begin{equation}
    d_\mathrm{SFR} = \frac{SFR_\mathrm{B19} - SFR_\mathrm{asloth}}{SFR_\mathrm{MD14} - SFR_\mathrm{B19}}\;.
\end{equation}
Then, we convert this maximum relative discrepancy into a goodness-of-fit parameter via
\begin{equation}
p_\mathrm{SFR} = \exp (-4d_\mathrm{SFR}) \;.
\end{equation}
This formula provides the correct limits of $p_\mathrm{SFR}=1$ for a good agreement between model and observation and an exponential penalty if the prediction of \asloth\ is outside the observed values. The factor of $4$ is an empirical weighting to give this observational constraint similar importance compared to the five observables.

\citet{dubois21} show with a 3D hydrodynamical simulation that the largest uncertainty to the SFRd is the cosmic variance in their zoomed-in simulated volume of approximately 4187 cMpc$^3$.  Our effective volume is smaller, which might even worsen this effect. \citet{Crosby13} find with their SAM that cosmic variance between different random realizations of the same cosmic volume can result in SFRd that differ by up to a factor of two. This is of the same order as the observational uncertainty of the SFRd at high redshift. We plan to investigate this variance in detail in a future study in a larger effective volume.

\subsection{Calibration Strategy}
Testing each new combination of parameters requires us to run \asloth\ using the 30 \ctp trees and the 8\,Mpc box, and takes about 1\,h wallclock time. Exploring the entire high-dimensional parameter space to obtain the full posterior distribution is prohibitively expensive for this optimization problem.

Therefore, we first identify three input parameters that are degenerate with other parameters and to which the observables are not very sensitive: $M_\mathrm{min}$, $\alpha _\mathrm{III}$, and $f_\mathrm{esc,III}$. The minimum mass of Pop~III stars is degenerate with the star formation efficiency, as a smaller value of $M_\mathrm{min}$ can be compensated with a larger value of $\eta _\mathrm{III}$. The escape fraction for Pop~III stars is degenerate with that for Pop~II stars, and the slope of the Pop~III IMF is also degenerate with the maximum mass of Pop~III stars and their star formation efficiency (SFE). Furthermore, we have verified that these three parameters have only a small effect on the fit of our six observables. We therefore keep them fixed during the calibration and will study their influence in more detail in a follow-up study.

The values of these three fixed parameters are chosen and motivated as follow. There are no observed metal-free stars in the MW, which constrains $M_\mathrm{min} \gtrsim 0.8 \Msun$ \citep{Hartwig15b,Ishiyama16,Magg19,Rossi2021}. On the other side, we have seen evidence of the chemical fingerprint of supernovae from metal-free progenitor stars with masses of $\sim 10 \Msun$ \citep{Ishigaki18}. We have chosen our fiducial $M_\mathrm{min}=5\Msun$ between those two limits.

A logarithmically flat IMF slope ($\alpha_\mathrm{III} = 1.0$) puts equal importance on each logarithmically spaced mass bin. This avoids any prior assumptions about dominant mass ranges. Moreover, several hydrodynamical simulations of Pop~III star formation yield results that are consistent with this IMF slope \citep{Greif11b,stacy16,wollenberg20,sharda21,latif22}.

The Pop~III SFRd in \asloth\ is rather high compared to other simulations at $z>20$ as we show below. In order to avoid that the Pop~III contribution to reionization requires an even higher SFRd, we set $f_\mathrm{esc, III}=60\%$, the highest value that is allowed by theory and observations \citep{dayal18}.

In Table~\ref{tab:parameters}, we list the six remaining free parameters of our model and the parameter ranges that we have explored while searching for an optimal model fit. We aim to fit six observables with six free parameters. Two of these observables (CSMF and SFRd) are functions (and not just single values) so that the final calibration problem is over-constrained.

\begin{deluxetable*}{llll}
\tablecaption{List of free parameters in \asloth. Top: six parameters that we calibrate based on observables. Bottom: three further parameters that we do not vary as we explain in the text. \label{tab:parameters}}
\tablewidth{0pt}
\tablehead{
\colhead{Parameter} & \colhead{Description} & \colhead{Explored Range} & \colhead{Best Fit} 
}
\decimalcolnumbers
\startdata
$M_\mathrm{max}$ & Maximum mass of Pop~III stars & $100-330$\Msun & $210$\Msun  \\
$\eta _{\rm III}$ & Pop~III star formation efficiency & $0.03-1.0$ & $0.38$ \\
$\eta _{\rm II}$ & Pop~II star formation efficiency & $0.01-1.0$ & $0.19$ \\
$\alpha_\mathrm{out}$ & slope of outflow efficiency & $0-2.5$ & $0.86$ \\
$M_\mathrm{out,norm}$ & normalization mass of outflow efficiency & $10^{9-11}\Msun$ & $7.5 \times 10^9$\Msun \\
$f_\mathrm{esc,II}$ & Pop~II ionizing photon escape fraction & $0-0.4$ & $0.37$ \\
\hline
$M_\mathrm{min}$ & Minimum mass of Pop~III stars & $5\Msun$ & $5\Msun$  \\
$\alpha_\mathrm{III}$ & slope of the Pop~III IMF & $1$ & $1$  \\
$f_\mathrm{esc,III}$ & Pop~III ionizing photon escape fraction & $0.6$ & $0.6$ \\
\enddata
\tablecomments{The explored range of $M_\mathrm{max}$ should enclose the mass range of PISNe and the upper limit for the escape fraction is based on \citet{dayal18}. Other explored ranges are based on previous experience \citep{chen22} and extensive experiments.}
\end{deluxetable*}

We expect further degeneracies of input parameters and we do not know if there is one unique global optimum, or several local optima instead. The goal of this optimisation strategy is therefore not to prove that we found the global optimum. Instead, we try to find a model configuration that reproduces all observables and that is robust to small variations in the input parameters.

We calibrate our model in two steps. First, we run a random parameter exploration in which we sample all input parameters randomly. During this step, we sample 128 different combinations of parameters. From these 128 parameter combinations, we choose the top four models that provide the highest value of $p_\mathrm{all}$. These four models are then used as the starting point of ``Quadient Descent'' (QD), a combination of quadratic fits and gradient descent that we have developed to calibrate \asloth.

Instead of optimizing all 6 input parameters at once, we calibrate \asloth\ with a sequence of one-dimensional optimizations. At each QD step, we choose a random input parameter that should be explored. Ideally, we want to calculate the gradient $\mathrm{d}p_\mathrm{all}/\mathrm{d}x_i$, where $x_i$ is the $i$-th input parameter that is currently under investigation. However, \asloth\ includes random sampling of variables in various places so that the resulting gradient $\mathrm{d}p_\mathrm{all}/\mathrm{d}x_i$ is not a continuous function of $x_i$.

\begin{figure}[ht]
\includegraphics[width=\columnwidth]{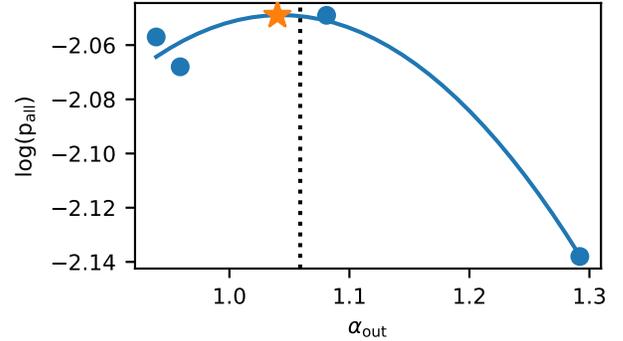}
\caption{Example of one QD step. We start at the previous value (vertical dotted line), sample 4 additional values (blue points), fit a quadratic function (blue line), and choose the maximum value as new parameter (orange star). \label{fig:QDstep}}
\end{figure}

Therefore, we cannot just sample one additional point to calculate the gradient. Instead, we calculate four additional models and fit a quadratic function to the obtained $p_\mathrm{all}(x)$. A quadratic fit is the lowest-order polynomial fit that is able to match local optima in the $p$-landscape and four sampled points are an efficient trade-off between sample density and computational time. The four explored values, $x_i$, of one parameter are chosen randomly based on a normal distribution centered at the previous value. Eventually, we choose the value of $x_i$ for which the fit predicts the highest $p_\mathrm{all}$ (inside the sampled parameter range). An example of one such QD step can be seen in Fig.~\ref{fig:QDstep}.
After several loops over all input parameters, the QD chains should converge to a similar solution. We stop these QD chains after around 350 iterations and analyze their explored and preferred parameter ranges in Fig.~\ref{fig:QDchains} in the Appendix.

All chains converge to a local optimum and we cannot find any combination of input parameters that provides a significantly better fit. Three out of four chains converge to the same optimal parameters. One chain prefers a slightly higher $\alpha_\mathrm{out}$. This higher value results in a better fit to the CSMF, but a worse fit to the MW stellar mass. The values for $p_\mathrm{all}$ for this chain are similar to the values for the other three chains.

To obtain one set of optimal parameters, we proceed in the following way. We cut the first 60 iterations, which guarantees that we only take sampled parameters into account for which the chains are already exploring near the optimum. Then, we take all the best fit parameters from all chains together and use the median value as our final best-fit value. Using the median is less sensitive to outlying chains.
The resulting best-fit values and observables of our best-fit model are summarized in Tab.~\ref{tab:fitted}. The best fitting parameters are reported in the last column of Tab.~\ref{tab:parameters}. While the Pop~III star formation efficiency of $\eta_\mathrm{III} = 0.38$ might appear high, we note that it is defined as the fraction of cold gas that forms stars per free-fall time. It is therefore not directly comparable to numerical simulations, which usually define the SFE as mass fraction and not per characteristic timescale \citep[e.g.][]{latif19}. We plan to explore the full posterior distributions of these parameters in a future study.

\subsection{Further Constraints}
\begin{figure}
\includegraphics[width=\columnwidth]{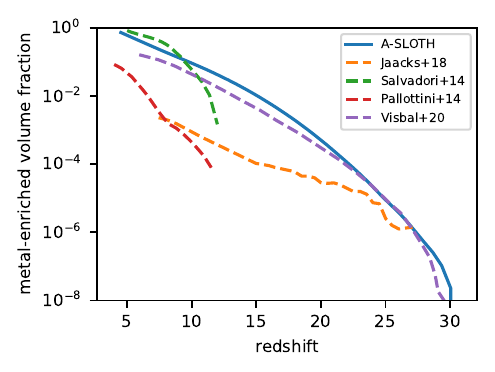}
\caption{\label{fig:Vmet}Metal-enriched volume fraction simulated with \asloth\ in the 8\,Mpc box (blue solid) compared to other results from the literature \citep{jaacks18b,salvadori14,pallottini14,visbal20}.}
\end{figure}

Besides these six observables that we actively try to reproduce, we also can compare our results to further results from the literature and from independent simulations. For example, we find a similar metal enrichment history as in other simulations (Fig.~\ref{fig:Vmet}).
Our results are in excellent agreement with \citet{visbal20} and in good agreement with other models.

\begin{figure}[ht]
\includegraphics[width=\columnwidth]{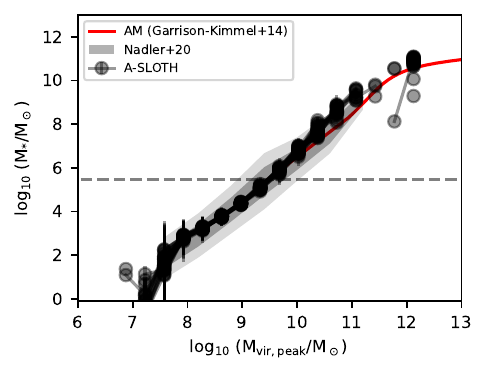}
\caption{We show the SMHMs at $z=0$ of 30 \ctp trees from our best model. At given \mpeak, we show the mean \mste (gray circles) and the standard deviation among the galaxies from each tree. We also plot the SMHM derived by \citet{Nadler2020} as gray contour, where the dark gray shows the 1\,$\sigma$ region and the light gray shows the 2\,$\sigma$ region. The SMHM derived by \citet{GarrisonKimmel} is plotted in red until the observational completeness. \label{fig:SMHM} }
\end{figure}
We show the SMHM relations produced by our fiducial model as gray curves in Fig.~\ref{fig:SMHM}. The circles show the mean stellar mass of galaxies at given $\mpeak$ along with 1$\sigma$ among them. The SMHM relation derived by \citet{GarrisonKimmel} is plotted in red, where the curve ends at $M_{*} = 2.9\times 10^5\Msun$ due to the observational completeness. We plot the SMHM relation of \citet{Nadler2020} as gray contours.

We find that the difference among SMHM relations from 30 \ctp trees is small. The larger scatter at $>10^{11} \Msun$ comes from the small statistics. The SMHM relations are consistent with both the relations presented by \citet{GarrisonKimmel} and \citet{Nadler2020} above the observational completeness and consistent with the relation of \citet{Nadler2020} below the observational completeness. We find a steepening of the SMHM relation below $\mpeak \approx 10^8 \Msun$. This is contrary to the finding in an earlier study \citep[that utilized an earlier version of \asloth]{chen22}, where they found a flattening of SMHM relation below $\mpeak \approx 10^9 \Msun$. The flattening presented by \citet{chen22} comes from the high Pop~II star formation efficiency (they adopted a fiducial value of $\eta_\mathrm{II} = 2$). This high $\eta_\mathrm{II}$ introduces a minimal stellar mass that forms before the stellar feedback kicks in and regulates further star formation. In our fiducial model, we have $\eta_\mathrm{II} = 0.19$, which avoids such a threshold. The physical models have also been improved, compared to the one used by \citet{chen22}, where the Pop~III star formation, stellar feedback, and the general chemical model were simpler than our fiducial model.

\begin{figure}[ht]
\includegraphics[width=\columnwidth]{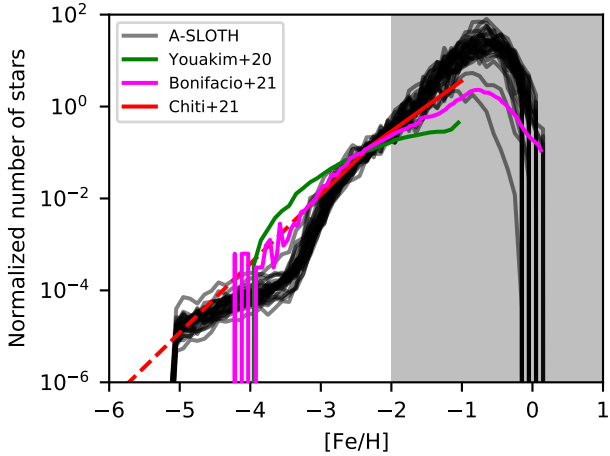}
\caption{We compare the MDF produced with the calibrated model and observation of \citet{Youakim20}, \citet{bonifacio21}, and \citet{chiti21}. The black lines show model results for all the Caterpillar trees. Other lines show three different observations. The histogram shows the number of stars in each bin. The sum of the histograms at $\mbox{[Fe/H]} < -2$ is normalized to be one for both model results and observations. \label{fig:MDF}}
\end{figure}

In Fig.~\ref{fig:MDF}, we show MDFs of the 30 \ctp trees and three observational estimates \cite[][]{Youakim20, bonifacio21, chiti21}. For low-metallicity stars, the distribution is reproduced well. For metal-rich stars ($\mbox{[Fe/H]} \gtrsim -2.5$), the contribution from type-Ia SNe is expected to be significant, therefore we do not claim that \asloth\ can reproduce the MDF also at higher metallicities.
Around [Fe/H] $=-3.5$, \asloth\ seems to systematically underproduce the observed MDF. This dip is related to the rather high Pop~III SFE in our calibrated model. However, a low SFE would not reproduce the fractions of EMP to UMP stars and of EMP to all stars. So this shape of the MDF is a compromise between other observables. We show below that the model with stochastic feedback is able to reproduce the shape of the MDF better.

\begin{figure}[ht]
\includegraphics[width=\columnwidth]{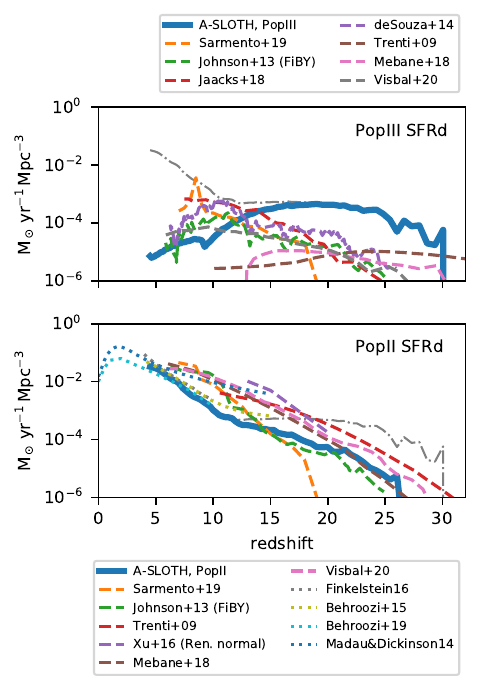}
\caption{Cosmic star formation rate densities for Pop~III stars (top panel) and Pop~II stars (bottom panel). The blue lines show our model prediction, the dashed lines show other models from the literature \citep{sarmento19,johnson13,jaacks18b,deSouza14,Trenti09,Mebane18,visbal20,Xu16b}, the dotted lines in the bottom panel show observations \citep{finkelstein16,BehrooziSilk2015,behroozi19,Madau14}, and the gray dashed-dotted lines in both panels show the total SFRd (Pop~II+Pop~III). \label{fig:SFRd}}
\end{figure}

We illustrate the cosmic star formation rate densities for Pop~III stars (top panel) and Pop~II stars (bottom panel) in Fig.~\ref{fig:SFRd}. Our Pop~III SFRd is higher at high redshift. However, when we compare the cumulative stellar mass density of Pop~III stars of \asloth\ with other models, our prediction is in the middle of other literature values. With \asloth\ we obtain a cumulative Pop~III stellar mass density of $9.5 \times 10^4 \Msun \mathrm{Mpc}^{-3}$. This is within the range of other models that provide predictions from $1.8 \times 10^3 \Msun \mathrm{Mpc}^{-3}$ \citep{Mebane18} to $2.9 \times 10^5 \Msun \mathrm{Mpc}^{-3}$ \citep{sarmento19}.

We also analyze the recovery time between Pop~III and Pop~II star formation. For this comparison, we define the recovery time as the time between the first Pop~III SN and the first Pop~II SN in a halo. This timescale is set by the radiative and mechanical feedback of the Pop~III star formation and the ability of the halo to retain or accrete new gas. We find recovery times in the range of $\sim 1\,\mathrm{Myr} - 1\,\mathrm{Gyr}$. This spread is in good agreement with numerical simulations \citep{Jeon14, Chiaki18, latif20, Hicks2021, Magg21a}. Although such a large range is seen in several different simulations, its cause is still poorly understood. \citet{Chiaki18} relate it to the ratio of the energy injected by feedback (from both photoionization and SNe) to the binding energy of the halo. In this picture, if the total energy output by a star is larger than the binding energy, the halo will undergo a long recovery process of at least several tens of millions of years. Alternatively, \citet{Magg21a} suggest it could be caused by variations in the amount of sub-structures present in the minihalos. These additional comparisons are further confirmation that our model for baryonic physics in minihalos works well.

\subsection{Calibration of EPS Mode}
To test EPS trees and our implementation of stochastic feedback, we compare three different cases. First, we run a \ctp tree with spatial feedback (our fiducial model). Then, we run exactly the same merger tree, but ignore the spatial information and use the stochastic feedback instead. Those two runs should produce similar results. Then, we run a third case in which we use an EPS-generated merger tree of a halo with the same mass as the MW with stochastic feedback. Ideally, the essential physics can be implemented appropriately, despite the difference in merger tree details, so that also this last case should produce similar results. However, with EPS, we follow only the formation history of the MW main halo and ignore subhalos and environmental effects. Therefore, we do not expect that this EPS-based run produces identical results to the \ctp-based runs. Previously, we have confirmed that external feedback affects the observables and that about 10\% of halos are enriched externally before they are enriched internally \citep{Tarumi20}.

\begin{figure}[htb]
\centering
\includegraphics[width=\columnwidth]{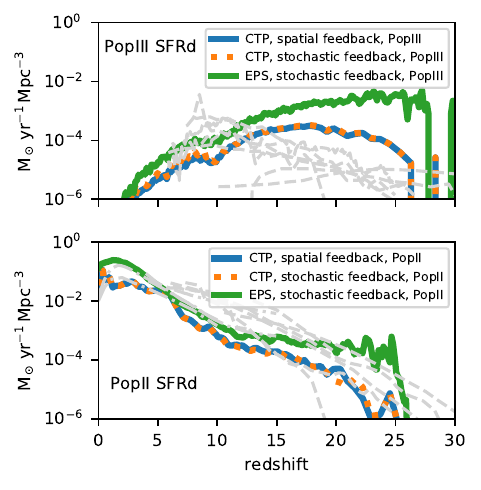}
\includegraphics[width=\columnwidth]{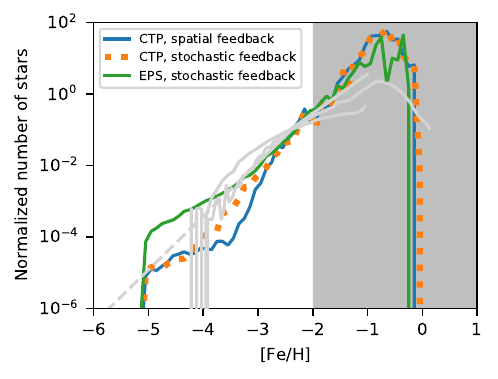}
\caption{Comparison of SFRd (top) and MDF (bottom) between different MW-like simulations. The gray lines are models from the literature (compare Fig.~\ref{fig:MDF} and Fig.~\ref{fig:SFRd}). We focus on the difference between our three models and the SFRd in a MW-like volume should not reproduce the cosmologically representative SFRd. Hence, we do not focus on the comparison to literature here, but rather provide them to guide the eye. The two models that are based on \ctp merger trees (blue, orange) provide similar results, whereas the EPS-based model differs at high redshift and low metallicity. \label{fig:EPS1}}
\end{figure}

\begin{figure}[htb]
\centering
\includegraphics[width=0.95\columnwidth]{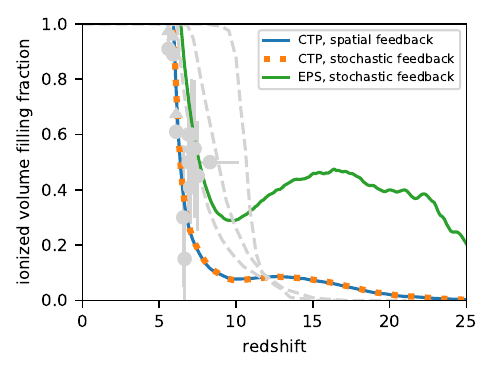}
\includegraphics[width=0.95\columnwidth]{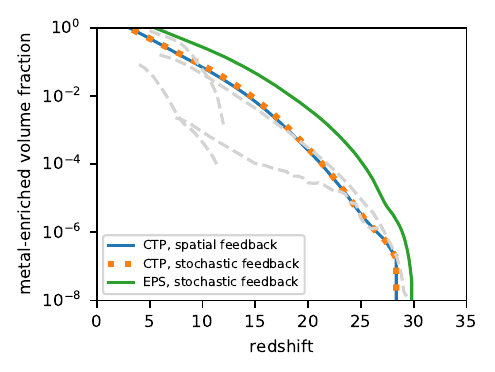}
\caption{Volume filling fraction of ionized regions (top) and metal-enriched regions (bottom). We compare three different approaches to model a MW-like galaxy with and without spatial information. The gray lines and symbols are models from the literature (compare Fig.~\ref{fig:tau} and Fig.~\ref{fig:Vmet}). These literature models should only guide the eye since a MW-like volume is not representative and these simulations do not intend to reproduce these literature values. The early peak of ionizing radiation in the EPS-based model is due to a smaller effective volume and due to a higher Pop~III SFRd. \label{fig:EPS2}}
\end{figure}

The comparison between different models of external feedback can be seen in Fig.~\ref{fig:EPS1} and Fig.~\ref{fig:EPS2}.
The two runs that are based on the same \ctp merger tree result in very similar star formation rates. The MDFs are slightly different because we do not accrete externally enriched metals in the run with stochastic feedback. The run based on EPS merger trees deviates from the spatially resolved merger trees at high redshift and low metallicities.

Fig.~\ref{fig:EPS2} shows how the volume filling fractions of ionized (top) and metal-enriched (bottom) regions evolve over time.
Again, the two runs based on \ctp merger trees are very similar. To obtain an estimate of the total volume, we divide the total mass of a merger tree at $z=0$ by the mean cosmic density. Since the MW is not cosmologically representative, we do not expect to reproduce the ionization history or cosmic evolution of metals in these test scenarios.

Overall, we see that our implementation of stochastic feedback is working. However, EPS trees are in general not very well suited to reproduce observables because of their simplified structure that cannot reproduce a MW-like galaxy with its satellites and cosmic environment. \asloth\ is calibrated and intended for the use with spatially resolved merger trees. We recommend EPS merger trees only for initial tests and the use of spatially resolved merger trees for all scientific applications.

\section{Summary}
\asloth\ is the first public SAM that was designed to simulate the formation of the first stars and galaxies. In this paper, we have shown how \asloth\ can reproduce six observables with our calibration strategy: the ionization history of the Universe, the stellar mass of the MW and its satellite galaxies, the relative fractions of EMP stars, and the SFRd at high redshift. 

We have shown that \asloth\ is computationally efficient and has versatile applications (Sec.~\ref{sec:PreviousSloth}). Users can start experimenting with the memory-friendly EPS mode, and then run scientific experiments on publicly available merger trees.
\asloth\ includes sophisticated models for external feedback (Sec.~\ref{sec:SpatialFeedback}), inhomogeneous metal mixing (Sec.~\ref{sec:mixing}), and baryonic physics (Sec.~\ref{sec:bc}) inside the first galaxies. All modules and parameters can be modified by the user in order to study new physics. It is the first SAM that samples the formation of individual Pop~III and Pop~II stars, which allows to model feedback by individual SNe in dwarf galaxies.

In addition to this code release paper, we also plan to publish the main data products of \asloth\ that can directly be used by the community, such as SFRd, recovery time distribution, and MW satellite properties with convenient fitting functions.

We also expect that \asloth\ can contribute to new scientific discoveries in the next decade, by optimizing surveys with JWST or interpreting LISA and third generation terrestrial detector observations. For example, \asloth\ can predict the optimal redshift to search for direct detections of Pop~III SNe \citep{magg16,Hartwig18b} or the contribution of Pop~III remnant binary black holes to the data stream of GW detectors. With these next-generation telescopes and surveys, we will greatly extend the frontier of observational cosmology, which is bound to result in serendipitous discoveries of unknown objects and events. Given its inherent adaptability and computational speed, \asloth\ is ideal for quickly interpreting such frontier observations, and helping to define follow-up campaigns.

\begin{acknowledgments}
We thank Boyuan Liu, Britton Smith, Anna Schauer, and Bet\"ul Uysal for valuable discussions. We also thank Stefania Salvadori, Rosa Valiante, and the anonymous referee for constructive feedback.
We acknowledge the work of \citet{LaceyCole1993} and \citet{Parkinson2008}, who developed \textsc{galform}, based on which we started the implementation of \asloth. We acknowledge funding from JSPS KAKENHI Grant Numbers 19K23437, 20K14464, and 20H05845. The team in Heidelberg is thankful for support from the Deutsche Forschungsgemeinschaft (DFG) via the Collaborative Research Center (SFB 881, Project-ID 138713538) ``The Milky Way System'' (sub-projects A1, B1, B2 and B8) and from the Heidelberg Cluster of Excellence (EXC 2181 - 390900948) ``STRUCTURES: A unifying approach to emergent phenomena in the physical world, mathematics, and complex data'', funded by the German Excellence Strategy. They also acknowledge funding form the European Research Council in the ERC Synergy Grant ``ECOGAL -- Understanding our Galactic ecosystem: From the disk of the Milky Way to the formation sites of stars and planets'' (project ID 855130). MV acknowledges funding from the European Research Council under the European Community's Seventh Framework Programme (FP7/2007-2013 Grant Agreement no.\ 614199, project ``BLACK''). The project benefited from computing resources provided by the State of Baden-W\"urttemberg through bwHPC and DFG through grant INST 35/1134-1 FUGG, and from the data storage facility SDS@hd supported through grant INST 35/1314-1 FUGG. We also acknowledge the Texas Advanced Computing Center (TACC) for providing HPC resources under XSEDE allocation TG-AST120024. And we acknowledge the Leibniz Computing Center (LRZ) for providing HPC resources in the project pr74nu.
\end{acknowledgments}


\software{\textsc{galform} \citep{LaceyCole1993,Parkinson2008}, python \citep{python09}, numpy \citep{harris20}, scipy \citep{virtanen20}, matplotlib \citep{hunter07}, pandas \citep{McKinney2010,Reback2022}.}



\appendix
\section{Supplementary Information on Method}
\subsection{Fitting Function of Halo Concentration}
\label{app:cdm}
We follow the fitting functions in \citet{Correa15} to compute the halo's concentration at $ 0 \leq z < 10$ and adopt a constant halo concentration of $3.5$ at $z > 10$. At $0 \leq z < 4$, we have:
\begin{equation}
\begin{aligned}
    &\mathrm{log}_{10} c_\mathrm{dm} = a + b\, \mathrm{log}_{10}(M_\mathrm{vir})\left(1+cM_\mathrm{vir}^2\right) \\
    &a = 1.7543 - 0.2766(1+z) + 0.02039(1+z)^2 \\
    &b = 0.2753 + 0.00351(1+z) - 0.3038(1+z)^{0.0269} \\
    &c = -0.01537 + 0.02102(1+z)^{-0.1475} \\
\end{aligned}
\end{equation}
and when $4 \leq z < 10$, we have:
\begin{equation}
\begin{aligned}
    &\mathrm{log}_{10} c_\mathrm{dm} = a + b\, \mathrm{log}_{10}(M_\mathrm{vir}) \\
    &a = 1.3081 - 0.1078(1+z) + 0.00398(1+z)^2 \\
    &b = 0.0223 - 0.0944(1+z)^{-0.3907}.
\end{aligned}
\end{equation}
At $z \geq 10$, the halo's concentration in the relevant virial mass range ($M_\mathrm{vir} > 10^5 \Msun$) has little variation (Fig.~7 in \citealt{Correa15}). Therefore, we adopt a constant concentration value and our results are insensitive to the adopted value.

\subsection{Alternative Models for the Critical Halo Mass}
\label{app:Mcrit}
The following two models for the critical halo mass of star formation are implemented as alternatives to our fiducial model \citep{Schauer21}. In model OHS, we compute $M_\mathrm{crit}$ by considering prescriptions in \citet{Oshea08}, \citet{Hummel12}, and \citet{Stacy11}. The idea is to combine models that take into account the LW background \citep{Oshea08} and baryonic streaming \citep{Stacy11} with our baseline model \citep{Hummel12} that we had used in previous versions of \asloth.
\citet{Oshea08} assumed that the critical mass of a halo depends solely on $J_{21}$. We denote this critical mass as $M_\mathrm{crit, O}$, given by
\begin{equation}
    M_\mathrm{ crit, O}/\Msun = 4\left( 1.25\times10^{5} + 8.7\times10^{5} \left( 4 \pi J_{21} \right)^{0.47} \right).
\end{equation} 
\citet{Hummel12} assumed that gas reaches high density and collapses to form Pop~III stars when a halo reaches a critical virial temperature, $T_\mathrm{ crit} = 2200$ K. The critical mass $M_\mathrm{ crit, H}$ is then dependent on redshift
\begin{equation}
    M_\mathrm{ crit, H}/\Msun = 10^6 \left( \frac{T_\mathrm{ crit}}{1000 \mathrm{ K}} \right)^{1.5} \left( \frac{1+z}{10} \right)^{-1.5}.
\end{equation}
Finally, we follow \citet{Stacy11} and compute a critical mass $M_\mathrm{ crit, S}$
\begin{equation}
    M_\mathrm{ crit, S}/\Msun = \frac{\pi v_\mathrm{ eff}^3}{6 G^{3/2} \rho^{1/2}}, 
\end{equation}
where $\rho$ is the mean DM density of the halo and $v_\mathrm{ eff} = \sqrt{v_{\rm BC}^2 \sigma_{\rm rms}(z)^{2} + c_\mathrm{s}^2}$ is the effective velocity of the gas. This depends on the sound speed in the IGM, $c_\mathrm{s} = \sqrt{k_\mathrm{ B}T/\mu m_\mathrm{ H}}$, where $T = 0.017(1+z)^2$~K prior to the onset of Pop III star formation \citep{Schneider15}, and on the redshift-dependent streaming velocity
\begin{equation}
    \frac{v_\mathrm{ BC} \sigma_{\rm rms}(z)}{\mathrm{cm}/\mathrm{s}} = v_\mathrm{ BC} \times \frac{6\times10^5}{201} \times (1+z), \label{eq:stream}
\end{equation}
where $v_\mathrm{BC}$ is the streaming velocity in units of the rms value.
    In model OHS, the combined critical mass is determined by taking the maximum of $M_\mathrm{ crit, O}$, $M_\mathrm{ crit, H}$, and $M_\mathrm{ crit, S}$, i.e.\
\begin{equation}
    M_\mathrm{ crit, OHS} = \mathrm{max}(M_\mathrm{ crit, O}, M_\mathrm{ crit, H}, M_\mathrm{ crit, S}).
\end{equation}

The third $M_\mathrm{ crit}$ model we implement in \asloth\ is based on the formula given in \citet{Fialkov13} and we denote this model as F13. This critical mass $M_\mathrm{ crit, F13}$ is computed by  
\begin{equation}
    \label{eq:f13_mcrit}
    M_\mathrm{ crit, F13}/\Msun = M_0( 1 + 6.96(4 \pi J_{21})^{0.47} ),
\end{equation}
where $J_{21}$ depends on redshift via $J_{21} = 10^{2-z/5}$, and $M_{0}$ is the critical mass when there is no LW background. 
We compute $M_0$ based on \citet{Fialkov12}: 
\begin{equation}
\label{eq:M_0}
    M_{0}/\Msun = \left( \frac{v_\mathrm{ cool}(z)}{146.6 \rm{km/s}} \right)^3 \times \Omega_\mathrm{ m, 0}^{-0.5} \times (1+z)^{-1.5} \times 10^{12} \mathrm{ h^{-1}},
\end{equation}
where $\Omega_\mathrm{ m, 0} = 0.3086$ is the matter density at $z=0$ and the redshift dependent circular velocity threshold $v_\mathrm{cool}(z)$ is given by 
\begin{equation}
    \frac{v_\mathrm{cool}(z)}{\mathrm{km}/\mathrm{s}} = \sqrt{ (3.714)^2 + (4.015 \times 10^{-5} v_\mathrm{ BC} \sigma_{\rm rms}(z))^2 },
\end{equation}
where $v_\mathrm{ BC} \sigma_{\rm rms}(z)$ is given by Equation~\ref{eq:stream} above. The full derivation of $M_0$ can be found in Appendix~D in \citet{chen22}.

In Fig.~\ref{fig:mcritmodel} we show the evolution of $M_\mathrm{crit}$ vs. redshift for several  different values of $v_\mathrm{ BC}$ for the three implemented models.
\begin{figure}
        \centering
        \includegraphics[width=0.5\linewidth]{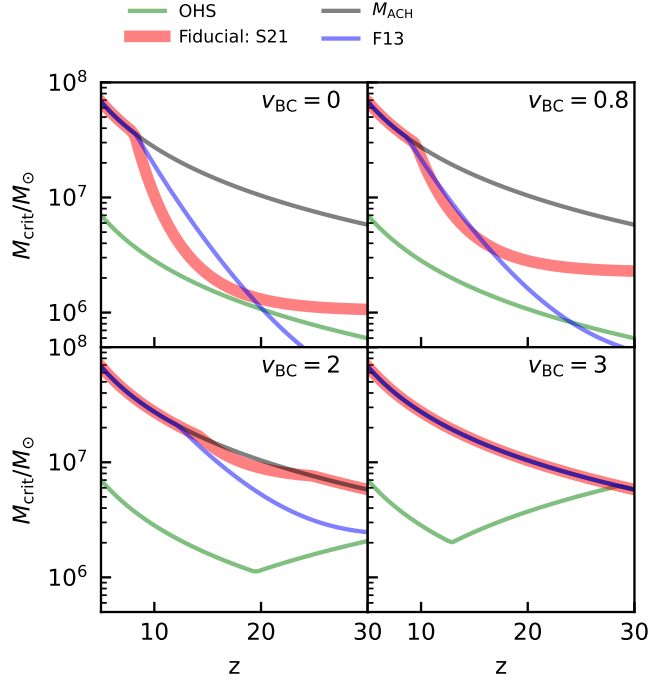}
        \caption{Redshift evolution of $M_\mathrm{crit}$ in our fiducial (S21) model and two other models, computed for several different values of the streaming velocity: $v_{\rm BC} = 0, 0.8, 2$ and 3 times $\sigma_{\rm rms}$. The values of $M_\mathrm{crit}$ in the OHS, S21, and F13 models are plotted in green, red, and blue, respectively. We additionally plot $M_\mathrm{ACH}$ in gray for reference.}
        \label{fig:mcritmodel}
\end{figure}

\subsection{Technical Implementation of Spatial Feedback}
\label{app:bubble}
Determining which ionized and enriched bubbles a halo sits inside of in its most simple implementation requires computing the distance between each possible pair of halos. This would lead to the computational cost of the simulations scaling as $N^2$ for $N$ halos and implies having to compute trillions of distances on each timestep. To avoid this we use a tree-based approach that for a given target halo limits the number of halos to which we need to compute the distances. 

\begin{figure}
\centering
    \includegraphics[width=0.5\linewidth]{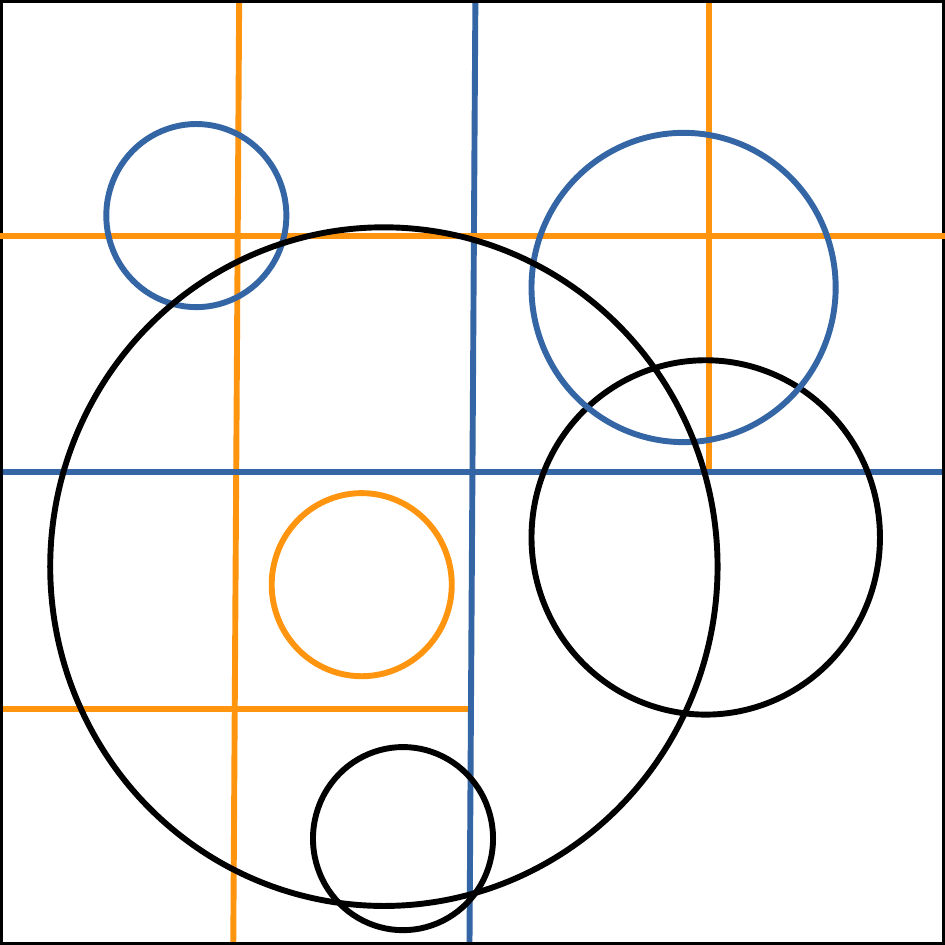}
    \caption{Illustration of the oct-tree construction for determining which bubbles a halo lies inside of. The colors represent the levels (black: 1st level, blue: 2nd level, orange: 3rd level). The bubbles (circles) are colored according to the node they are associated with. Each bubble is associated with the smallest tree node it is completely contained within.}
    \label{fig:tree}
\end{figure}

In this approach we build an oct-tree of bubbles and consecutively add all bubbles during a timestep to it. The tree is started with a head-node encompassing the entire physical volume. This node is split into 8 sub-nodes (i.e., it is split in two along each dimension) and the sub-nodes are split into 8 smaller nodes. We limit the tree to a depth of 20 levels, but only allocate nodes that actually contain at least one bubble. Each bubble is associated with exactly one node of this oct-tree, namely the smallest node that fully encompasses the bubble. Thus, large bubbles, and bubbles that are very close to edges of the larger nodes are located in the first few levels, small bubbles are located in the lower levels. This is illustrated in Fig. \ref{fig:tree}.

In order to find all bubbles that overlap with a point (e.g., the center of a halo), one only has to test bubbles associated with tree-nodes the point is inside of. This algorithm results in identical results as testing all pairwise combinations of halos at a substantially reduced computational cost.

\subsection{Fitting Functions for Inhomogeneous Metal Mixing}
\label{sec:dZfit}
In \citet{Tarumi20}, we obtain fitting functions for internal and external enrichment separately. For internal enrichment, we use a Gaussian distribution function with $\mu = -0.03, \sigma = 0.15$:
    \begin{equation}
        p(x) = \frac{1}{\sqrt{2\pi (0.15)^{2}}}\exp\biggl[\frac{(x+0.03)^2}{2 (0.15)^{2}}\biggr].
        \label{eq:Gaussian}
    \end{equation}
For external enrichment, we use an exponentially modified Gaussian distributions. The distribution has three free parameters, $(K, \mu, \sigma)$, and the probability distribution function $p(x; \mu, \sigma, \lambda)$ is:
    \begin{equation}
        p(x; \mu, \sigma, \lambda)=
\frac{\lambda}{2} \exp\biggl[\frac{\lambda}{2}(2\mu+\lambda\sigma^{2}-2x)\biggr]{\rm erfc}\biggl(\frac{\mu+\lambda\sigma^2-x}{\sqrt{2}\sigma}\biggr), \\
    \label{eq:PdZ_external}
    \end{equation}
    where
    \begin{eqnarray}
        {\rm erfc}(x) 
        &=& 1 - {\rm erf}(x)\\
        &=& \frac{2}{\sqrt{\pi}}\int^{\infty}_{x}e^{-t^{2}}dt,
    \label{eq:erfc}
    \end{eqnarray}
and the three parameters are given in Table.~\ref{tab:fitting_result}.
    
\begin{table}
	\centering
	\caption{Parameters to be used for the external enrichment fitting function. These parameter sets are for ``$-dZ$'', not $dZ$ itself.} 
	\label{tab:fitting_result}
	\begin{tabular}{l|lll} 
		\hline
		Metallicity & $\mu$ & $\lambda$ & $\sigma$ \\
		\hline
		$-2<Z\leq -1$ & -0.07 & 0.14 & 2.78\\
		$-3<Z\leq -2$ & -0.01 & 0.24 & 1.41\\
		$-4<Z\leq -3$ & 0.08 & 0.20 & 0.77\\
		$-5<Z\leq -4$ & 0.16 & 0.19 & 0.38\\
		$-8<Z\leq -5$ & 1.63 & 1.01 & 0.35\\
	\end{tabular}
\end{table}

\section{Supplementary Information on Calibration}

\subsection{Random Parameter Exploration}
\label{sec:AppPara}
\begin{figure}[ht]
\centering
\includegraphics[width=0.85\textwidth]{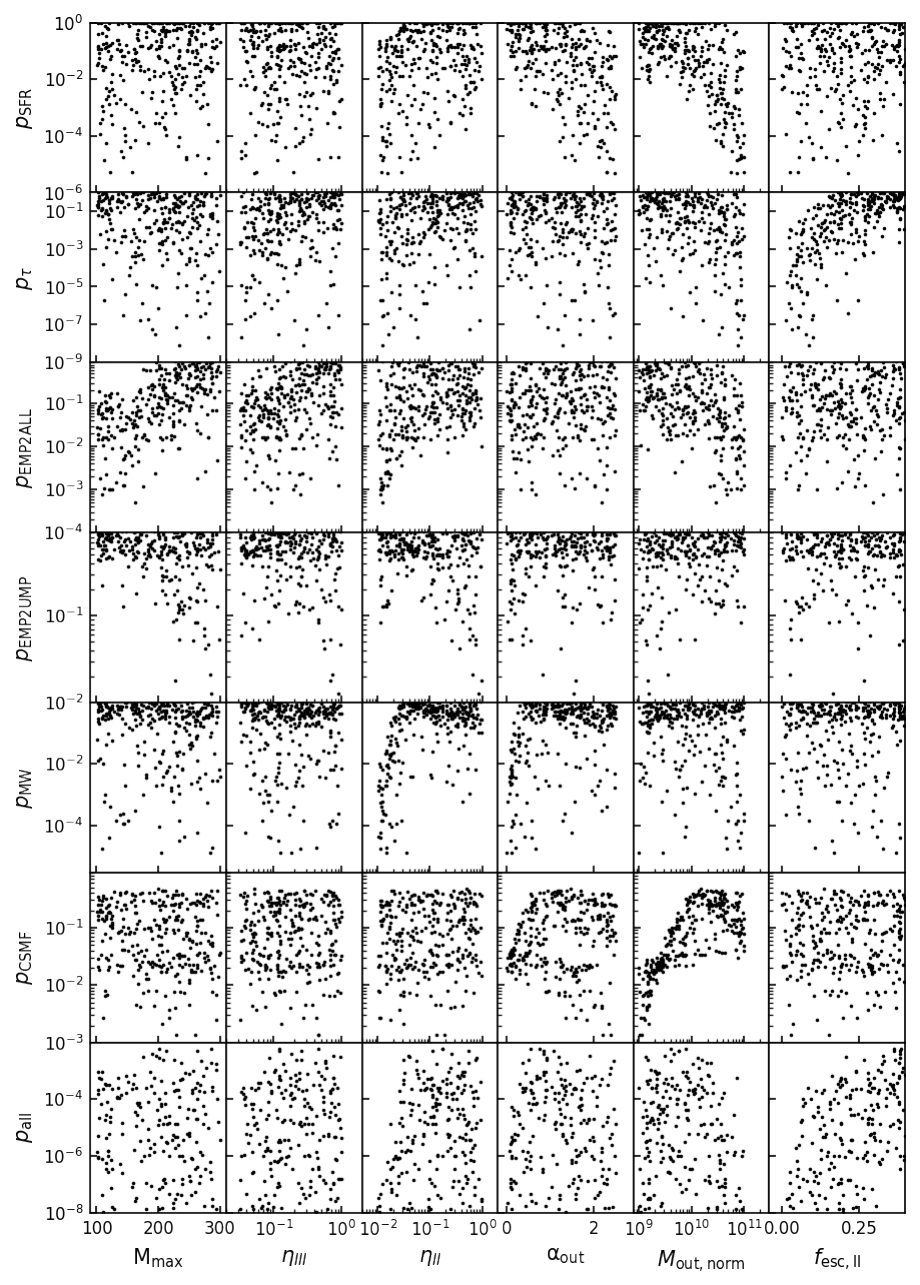}
\caption{Correlations between the input parameters (horizontal) and goodness-of-fit parameters (vertical) for 128 randomly chosen combinations of input parameters. This plot illustrates which observables are sensitive to which input parameters.\label{fig:pGrid}}
\end{figure}

The first step of our calibration is that we run 128 different models with random input parameter combinations. In Fig.\ref{fig:pGrid} we show how the goodness-of-fit parameters depend on the input parameters.
For example, the optical depth is sensitive to the ionizing escape fraction and the CSMF of MW satellites is sensitive to our normalization mass of supernova feedback. Furthermore, the maximum mass and SFE of Pop~III stars is sensitive to EMP-based quantities. In the last row ($p_\mathrm{all}$) we see which parameter ranges are preferred or disfavored. For example, models with $M_\mathrm{max}>200\Msun$ can reach higher values of $p_\mathrm{all}$ than models with $M_\mathrm{max}<200\Msun$. Overall, this representation demonstrates that such a random parameter exploration in this 6D space is not effective to find a global optimum. Hence, we only use it as a first exploration of the parameter space and then optimize the calibration with QD.

\subsection{Quadient Descent}
\begin{figure}[ht]
\plottwo{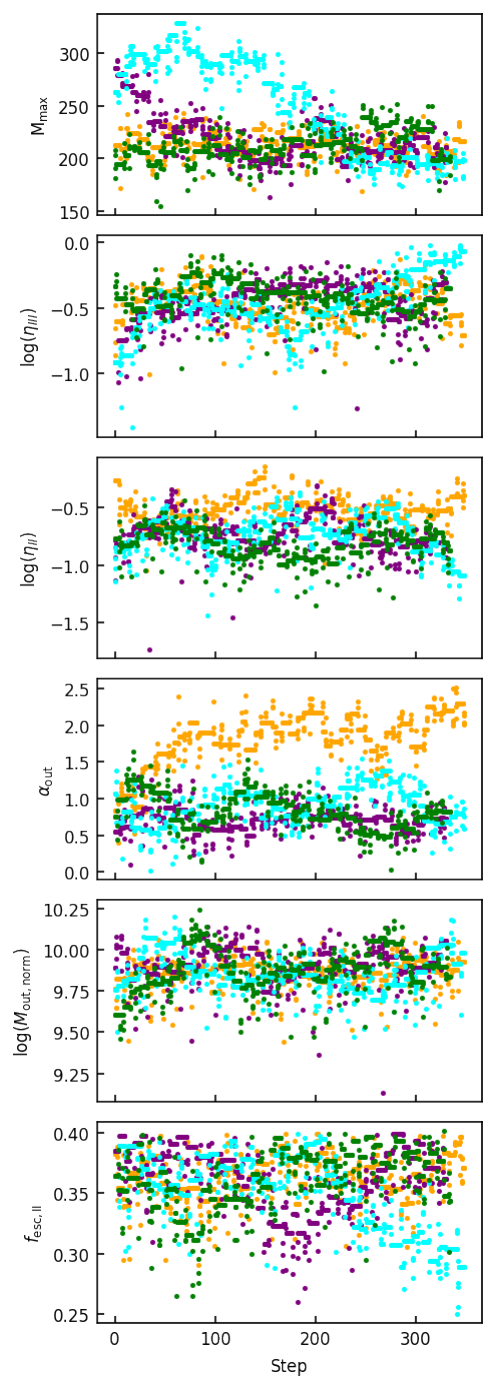}{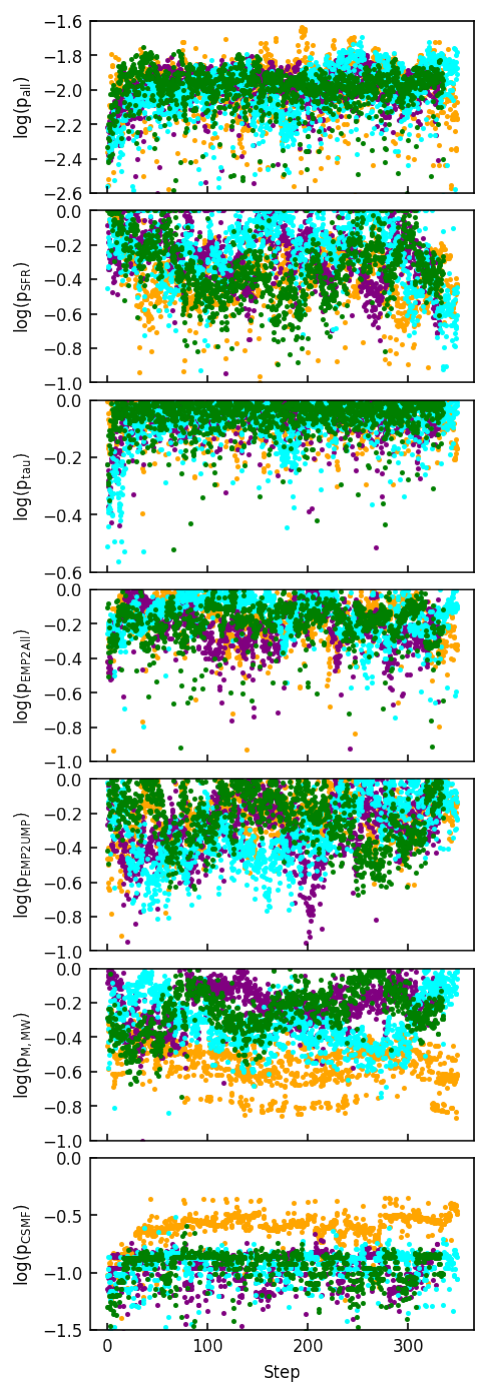}
\caption{Iterative exploration and optimization of the input parameters (left) and how the goodness-of-fit parameter change during this optimization (right).\label{fig:QDchains}}
\end{figure}
In Fig.~\ref{fig:QDchains}, the four colors show four independent chains. Each point shows one explored model and each step consists of 4 different models. The four chains converge to optimal values for all input parameter. Only the orange chain prefers a slightly higher value for $\alpha _\mathrm{out}$. As we can see in the bottom two panels, this higher value of the feedback efficiency exponent yields a better fit to the CSMF, but a worse fit to the MW stellar mass. This is a legitimate trade-off between two individual goodness-of-fit parameters, and $p_\mathrm{all}$ converges to similar values for all four chains.

\bibliography{MyBib}{}
\bibliographystyle{aasjournal}

\end{document}